\documentclass[useAMS,usenatbib]{mn2e}

\usepackage{graphicx,bm,amsmath,multirow,subfigure,longtable,float,color}

\usepackage[T1]{fontenc}
\usepackage{lmodern}
\usepackage{amsfonts,dsfont,upgreek}
\newcommand{\Var}{\mbox{$\mathrm{Var}$}}

\newcommand{\Cov}{\mbox{$\mathrm{Cov}$}}

\title[Weak lensing by galaxy troughs in DES SV]{Weak lensing by galaxy troughs\\ in DES Science Verification data}
\author[Gruen et al.]
{\parbox{\textwidth}{
D.~Gruen$^{1,2\star}$, O.~Friedrich$^{1,2}$, A.~Amara$^{3}$, D.~Bacon$^{4}$, C.~Bonnett$^{5}$, W.~Hartley$^{3}$, B.~Jain$^{6}$, M.~Jarvis$^{6}$, T.~Kacprzak$^{3}$, E.~Krause$^{7}$, A.~Mana$^{1,2}$, E.~Rozo$^{8}$, E.~S.~Rykoff$^{7,9}$, S.~Seitz$^{1,2}$, E.~Sheldon$^{10}$, M.~A.~Troxel$^{11}$, V.~Vikram$^{12}$, T. M. C.~Abbott$^{13}$, F.~B.~Abdalla$^{14,15}$, S.~Allam$^{16}$, R.~Armstrong$^{17}$, M.~Banerji$^{18,19}$, A.~H.~Bauer$^{20}$, M.~R.~Becker$^{7,21}$, A.~Benoit-L{\'e}vy$^{14}$, G.~M.~Bernstein$^{6}$, R.~A.~Bernstein$^{22}$, E.~Bertin$^{23,24}$, S.~L.~Bridle$^{11}$, D.~Brooks$^{14}$, E.~Buckley-Geer$^{16}$, D.~L.~Burke$^{7,9}$, D.~Capozzi$^{4}$, A.~Carnero~Rosell$^{25,26}$, M.~Carrasco~Kind$^{27,28}$, J.~Carretero$^{5,20}$, M.~Crocce$^{20}$, C.~E.~Cunha$^{7}$, C.~B.~D'Andrea$^{4,29}$, L.~N.~da Costa$^{25,26}$, D.~L.~DePoy$^{30}$, S.~Desai$^{31,32}$, H.~T.~Diehl$^{16}$, J.~P.~Dietrich$^{31,32}$, P.~Doel$^{14}$, T.~F.~Eifler$^{6,33}$, A.~Fausti Neto$^{25}$, E.~Fernandez$^{5}$, B.~Flaugher$^{16}$, P.~Fosalba$^{20}$, J.~Frieman$^{16,34}$, D.~W.~Gerdes$^{35}$, R.~A.~Gruendl$^{27,28}$, G.~Gutierrez$^{16}$, K.~Honscheid$^{36,37}$, D.~J.~James$^{13}$, K.~Kuehn$^{38}$, N.~Kuropatkin$^{16}$, O.~Lahav$^{14}$, T.~S.~Li$^{30}$, M.~Lima$^{25,39}$, M.~A.~G.~Maia$^{25,26}$, M.~March$^{6}$, P.~Martini$^{36,40}$, P.~Melchior$^{36,37}$, C.~J.~Miller$^{35,41}$, R.~Miquel$^{5,42}$, J.~J.~Mohr$^{2,31,32}$, B.~Nord$^{16}$, R.~Ogando$^{25,26}$, A.~A.~Plazas$^{33}$, K.~Reil$^{9}$, A.~K.~Romer$^{43}$, A.~Roodman$^{7,9}$, M.~Sako$^{6}$, E.~Sanchez$^{44}$, V.~Scarpine$^{16}$, M.~Schubnell$^{35}$, I.~Sevilla-Noarbe$^{27,44}$, R.~C.~Smith$^{13}$, M.~Soares-Santos$^{16}$, F.~Sobreira$^{16,25}$, E.~Suchyta$^{36,37}$, M.~E.~C.~Swanson$^{28}$, G.~Tarle$^{35}$, J.~Thaler$^{45}$, D.~Thomas$^{4,46}$, A.~R.~Walker$^{13}$, R.~H.~Wechsler$^{7,9,21}$, J.~Weller$^{1,2,32}$, Y.~Zhang$^{35}$, and J.~Zuntz$^{11}$}
\vspace{0.4cm}\\
\parbox{\textwidth}{Author affiliations are listed at the end of this paper\\ $\star$ E-mail: \texttt{\rm \texttt{dgruen@usm.uni-muenchen.de}}}}

\begin{document}

\date{}

\pagerange{\pageref{firstpage}--\pageref{lastpage}} \pubyear{2015}

\maketitle

\label{firstpage}

\begin{abstract}
We measure the weak lensing shear around galaxy troughs, i.e.~the radial alignment of background galaxies relative to underdensities in projections of the foreground galaxy field over a wide range of redshift in Science Verification data from the Dark Energy Survey. Our detection of the shear signal is highly significant ($10$-$15\sigma$ for the smallest angular scales) for troughs with the redshift range $z\in[0.2,0.5]$ of the projected galaxy field and angular diameters of 10~arcmin $\ldots1^{\circ}$. These measurements probe the connection between the galaxy, matter density, and convergence fields. By assuming galaxies are biased tracers of the matter density with Poissonian noise, we find agreement of our measurements with predictions in a fiducial $\Lambda$ cold dark matter model. The prediction for the lensing signal on large trough scales is virtually independent of the details of the underlying model for the connection of galaxies and matter. Our comparison of the shear around troughs with that around cylinders with large galaxy counts is consistent with a symmetry between galaxy and matter over- and underdensities. In addition, we measure the two-point angular correlation of troughs with galaxies which, in contrast to the lensing signal, is sensitive to galaxy bias on all scales. The lensing signal of troughs and their clustering with galaxies is therefore a promising probe of the statistical properties of matter underdensities and their connection to the galaxy field.
\end{abstract}


\section{Introduction}

The measurement of weak gravitational lensing probes matter inhomogeneities in the Universe by means of the differential deflection they induce on the light of background sources. Most lensing analyses are driven by the signatures of matter overdensities, e.g.~the gravitational shear of galaxies \citep[e.g.][]{1996ApJ...466..623B,2004ApJ...606...67H,2004AJ....127.2544S,2006MNRAS.368..715M,2011A&A...534A..14V,2013MNRAS.432.1046B,2014MNRAS.437.2111V,desggl} and clusters of galaxies \citep[e.g.][]{1990ApJ...349L...1T,2009ApJ...701L.114M,2009ApJ...703.2217S,2012ApJ...754..119M,2012MNRAS.427.1298H,2012arXiv1208.0597V,wiscy,2014ApJ...795..163U} or the spatial correlation of shear due to intervening large-scale structure, cosmic shear \citep[e.g.][]{2000Natur.405..143W,2008A&A...479....9F,2010A&A...516A..63S,2013MNRAS.430.2200K,descs}.

Probes of underdense structures are complementary to this. Differences between dark energy and modified gravity (MG) models for cosmic acceleration might be more easily differentiable in cosmic voids \citep{2013MNRAS.431..749C,2014arXiv1408.5338L,2014arXiv1410.1510C}. The reason for this is that the screening of the hypothetical fifth force of MG \citep{vainshtein,2004PhRvD..69d4026K}, required to meet Solar system constraints, is absent in these low-density environments. MG therefore entails that negative density perturbations should grow more rapidly than predicted by General Relativity, with effects on the density profile in and around such structures \citep[cf.][]{2014arXiv1410.1510C}.

Using voids detected in Sloan Digital Sky Survey (SDSS) spectroscopic galaxy catalogues \citep{2012ApJ...761...44S,2014MNRAS.442.3127S,2014arXiv1410.0355L}, first measurements of the radial alignment of background galaxies have been made \citep{2014MNRAS.440.2922M,2014arXiv1404.1834C}. Future spectroscopic surveys will yield lensing measurements of void matter profiles with moderate signal-to-noise ratio (SNR; \citealt{2013ApJ...762L..20K}). Combined with predictions for void profiles \citep{2014PhRvL.112y1302H,2014PhRvL.112d1304H}, these will provide unique tests of gravity.

On large enough scales, we expect a symmetry between the excess and deficit of matter relative to the mean. Despite this fact, highly significant lensing measurements have thus far only been performed on matter overdensities. In this work, we follow the new approach of measuring the properties of underdense regions in projections of the galaxy density field over wide ranges in the radial coordinate, i.e.~in redshift. Due to the wide redshift range (e.g.~$z\in[0.2,0.5]$) used for the projection, these can be straightforwardly identified in galaxy catalogues with photometric redshift (photo-$z$) estimates. Because the selection from the projected field avoids underdensities which are randomly aligned with massive structures in front or behind them along their lines of sight, the SNR of the lensing signal is comparatively high. The approach is related to and motivated by the measurement of shear around galaxies in underdense projected environments, for which \citet{2013MNRAS.432.1046B} have previously detected significant radial aligment (cf.~their fig.~25; see also \citealt{gillis} for a detection of radial shear around galaxies in underdense 3D environments).

We make these measurements using Science Verification (SV) data from the Dark Energy Survey (DES, \citealt{2005astro.ph.10346T,2005IJMPA..20.3121F}). The data was taken after the commissioning of the Dark Energy Camera (DECam; \citealt{2008SPIE.7014E..0ED}; \citealt{2015arXiv150402900F}) on the 4m Blanco telescope at the Cerro Tololo Inter-American Observatory (CTIO) in Chile to ensure the data quality necessary for DES. We make use of a contiguous area of 139~deg$^2$ for which SV imaging data at lensing quality is available (cf.~also \citealt{2015arXiv150403002V,shapecat}).

The structure of this paper is as follows. In Section~2, we describe the data used. Our theoretical modelling of the trough signal is introduced in Section~3. Section~4 presents our measurements of the shear signal around troughs and their angular two-point correlation with galaxies.  We summarize and give an outlook to future work in Section~5. For all theory calculations, we use a fiducial flat $\Lambda$ cold dark matter ($\Lambda$CDM) cosmology ($\Omega_{\rm m}=0.3$, $\Omega_{\rm b}=0.044$, $\sigma_8=0.79$, $n_{\rm s}=0.96$). 

\section{Data}

We select trough positions and measure their lensing signal with galaxy catalogues from the SV phase of DES. In this section, we briefly describe the catalogues used.

\subsection{Galaxy catalogue}
\label{sec:galaxies}

The DES SVA1 red-sequence Matched Filter Galaxy Catalog (redMaGiC;
\citealt{redmagic}) is a photometrically selected luminous red galaxy (LRG)
sample chosen to have precise and accurate photometric redshifts.
redMaGiC makes use of the red sequence model computed from the
redMaPPer cluster catalogue (\citealt{2014ApJ...785..104R}; \citealt{redmapper}).  This model of the red sequence as a function of magnitude and
redshift is used to compute the best-fitting photo-$z$ for all galaxies
regardless of SED type, as well as the $\chi^2$ goodness-of-fit.  At any
given redshift, all galaxies fainter than a minimum luminosity
threshold $L_{\mathrm{min}}$ are rejected.  In addition, redMaGiC
applies a $\chi^2$ cut such that $\chi^2<\chi^2_{\mathrm{max}}$, where
the maximum $\chi_{\rm max}^2(z)$ is chosen to ensure that the resulting galaxy
sample has a nearly constant comoving space density $\bar{n}$.  In this work,
we use the \emph{high density} sample, such that $\bar{n} =
10^{-3}h^3\mathrm{Mpc}^{-3}$ and $L_{\mathrm{min}}(z) = 0.5L_{\star}(z)$ in the assumed cosmology.   This space density is
roughly four times that of the SDSS BOSS CMASS sample.  As detailed in
\citet{redmagic}, the redMaGiC photo-$z$s are nearly unbiased,
with a scatter of $\sigma_z/(1+z) \approx 1.7\%$ and a 4$\sigma$ outlier rate of about 1.7\%. By virtue of this, redshift errors are negligible for the selection of underdensities in the galaxy field when the latter is projected over the wide redshift ranges we use (see Section~\ref{sec:troughselection}).

\subsection{Trough selection}
\label{sec:troughselection}

The troughs are selected as centres of cylindrical regions (or, more accurately, of conical frustums) of low galaxy density as follows. Let the galaxy catalogue (cf.~Section~\ref{sec:galaxies}) be given with entries $\bm{x}_i,z_i,L_i$ for the angular position, redshift and luminosity of galaxy $i$, $1\leq i \leq n$. 

Define a function $W_z(z,L)$ that assigns a weight to each galaxy based on its redshift and luminosity. Furthermore, define $W_{\rm T}(\theta)$ to weight points by their projected angular separation $\theta$ from the centre. From these we define a projected, weighted and smoothed version of the galaxy density field 
\begin{equation}
G(\bm{x})=\sum_{i=1}^{n} W_{\rm T}(|\bm{x}-\bm{x}_i|) W_z(z_i,L_i) \; .
\end{equation}

In this study, we use a simple weighting corresponding to a hard cut in luminosity, redshift and radius, i.e.
\begin{align}
W_z(z,L)&= \left\lbrace\begin{array}{rl}1\;,& L\geq0.5L^{\star} \; \wedge \; 0.2\leq z\leq0.5 \\ 0\;,&\mathrm{otherwise}\end{array}\right. \nonumber \\
W_{\rm T}(\theta)&= \left\lbrace\begin{array}{rl}1\;,&  \theta\leq\theta_{\rm T} \\ 0\;,&\mathrm{otherwise}\end{array}\right.
\end{align}
with some trough radius $\theta_{\rm T}$, which we vary between 5 and 30~arcmin. The redshift range of $z\in[0.2,0.5]$ is motivated by a trade-off between having a large trough volume (and hence a large signal) and having sufficient background galaxies with useful shear estimates to measure the effect, although we test other settings in Section~\ref{sec:tomo}. Note that a luminosity-independent weighting scheme of LRGs, as we apply it here, is not far from optimal for the reconstruction of the matter field \citep{2011MNRAS.412..995C}.

\begin{figure}
\centering
\includegraphics[width=0.48\textwidth]{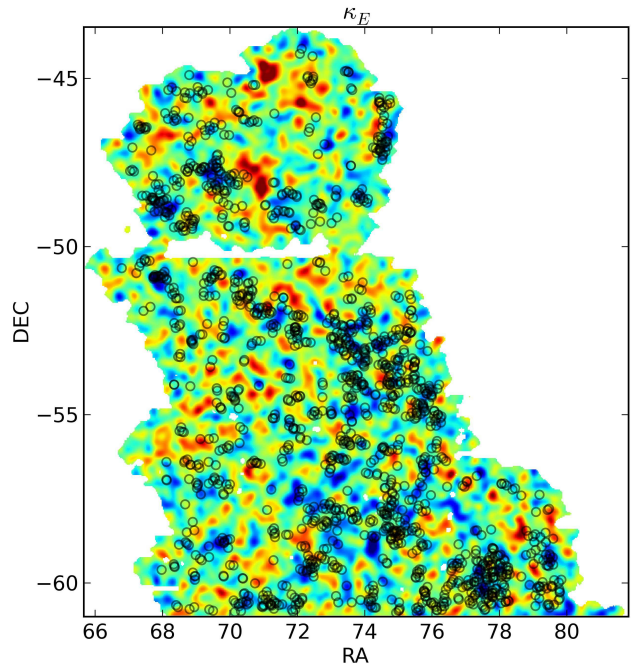}
\caption{Positions of 10~arcmin troughs (black circles, to scale, randomly selected sample of 1500 out of the $\approx$110,000 trough positions probed) overlaid on to lensing convergence map (red: positive, blue: negative). Convergence was estimated as described in \citet{2015arXiv150403002V} with a weighted lensing source catalogue (Section~\ref{sec:sourcecat}) and a 7~arcmin Gaussian smoothing.}
\label{fig:map}
\end{figure}

Probed on a finely spaced grid of sky positions, for which we use an $N_{\rm side}=4096$ \textsc{healpix} \citep{2005ApJ...622..759G} map with 0.86~arcmin pixel spacing, there is a distribution of weighted galaxy counts as measured by $G$. We select the set of trough positions $T$ as the points below the 20th percentile $G_{20}$ of that distribution, i.e. $T=\lbrace \bm{x}:G(\bm{x})\leq G_{20}\rbrace$.\footnote{We have also tried stricter (i.e.~lower percentile) and looser (i.e.~higher percentile) thresholds, which yield a higher (lower) amplitude of the signal with larger (smaller) uncertainties. Also see Section~\ref{sec:symmetry} for the selection of \emph{high}-density cylinders.} It is important to note that we are \emph{not} selecting individual minima of the galaxy density field. Instead, troughs are overlapping within and between samples of fixed trough radius. A map of trough positions for $\theta_{\rm T}=10$~arcmin is shown in Fig.~\ref{fig:map}. Note how the troughs visibly trace underdensities in the convergence field.

In practice, $G$ needs to be corrected for the effect of masking due to survey boundaries or bright stars, for example. A homogeneous masking fraction over all cylinders simply decreases the tracer density, which is accounted for by our model automatically. When masking fractions are not homogeneous but vary among the cylinders, the situation is more complicated. For the present analysis, we make the approximation of excluding all cylinders where more than $f=20$\% of the area inside the trough radius is masked in the galaxy catalogue and assume that all remaining cylinders have equal masking fractions when selecting troughs and modelling the signal. At the level of statistical precision achieved here, this simplification is not expected to cause a significant difference.

The mean surface density of redMaGiC galaxies in the useable area is approximately $\bar{n}=0.055$~arcmin$^{-2}$, corresponding to a mean count of approximately 4, 17, 69 and 155 galaxies in cylinders of radius 5, 10, 20 and 30~arcmin. At the lower 20th percentile, selected troughs have mean counts of 1, 9, 44 and 108 galaxies, respectively.

\subsection{Lensing source catalogue}
\label{sec:sourcecat}

For the background sources, we use a shape catalogue measured with \textsc{ngmix}.\footnote{cf.~\texttt{https://github.com/esheldon/ngmix}} We apply the cuts, weighting and responsivity correction as recommended in \citet{shapecat}, where also the shape measurement and testing of catalogues is described in detail. In order to prevent confirmation bias, shear estimates in the catalogue were blinded with an unknown factor until the analysis had been finalized (cf. \citealt{shapecat}, their section 7.5).

\begin{figure}
\centering
\includegraphics[width=0.48\textwidth]{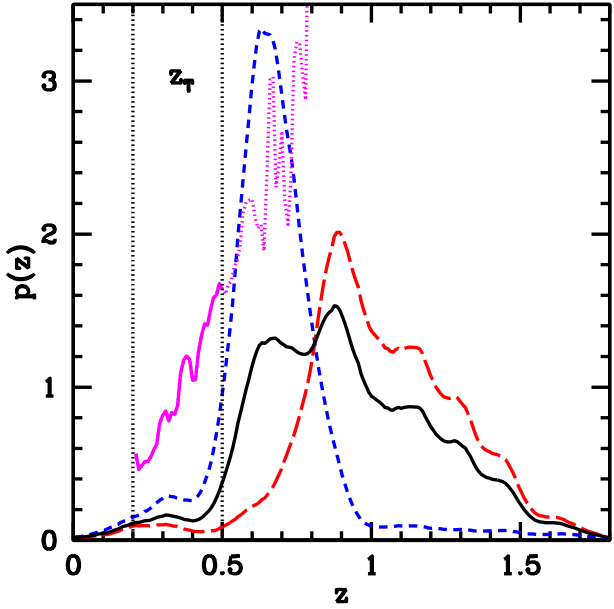}
\caption{Redshift distributions of the low-redshift lensing source sample (blue, short dashed line), high-redshift sample (red, long dashed line) and the combined fiducial source sample (black, solid line). All distributions are weighted by inverse shape noise of the sources in analogy to the shear signal and normalized for $\int p(z)\;\mathrm{d}z=1$. Redshift distribution of all redMaGiC galaxies is shown as the magenta dotted line (solid for the fiducial trough redshift range $z\in0.2,\ldots,0.5$ indicated by the dotted vertical lines).}
\label{fig:z}
\end{figure}

We use the two highest redshift bins defined in \citet[][cf.~their fig.~3]{descs} by means of photo-$z$ probability density estimates made with \textsc{SkyNet} \citep{2013ascl.soft12007G,2015MNRAS.449.1043B}, a method that performed well in an extensive set of tests on SV data \citep{2014MNRAS.445.1482S}. The mean redshift of the lower (higher) redshift bin is $z\approx0.6$ ($z\approx0.9$). We use the appropriately weighted \textsc{SkyNet} stacked $p(z)$ estimate for predicting the lensing signal (cf. Section~\ref{sec:model}). To maximize the SNR for our non-tomographic measurements, we weight the signal measured in both bins as 1:2 to approximately accommodate the ratio of effective inverse critical surface mass density. The resulting $p(z)$ of the samples used are shown in Fig.~\ref{fig:z}. The tests performed in \citet{photoz} indicate that errors in photo-$z$ estimation are not a dominant systematic error for the prediction of the shear power spectrum that we use them for.

We note that the consistency of different shape measurement and photo-$z$ methods with the catalogues used in this analysis has been investigated in detail and confirmed within the systematic requirements on the present data in several works \citep{2014MNRAS.445.1482S,shapecat,photoz,descs,descsc}.

\section{Theory}
\label{sec:model}

Structures in the Universe can be described by an underlying field, the matter density as a function of position and time. Matter density itself is not an observable. Its properties can, however, be recovered by a number of observable fields, such as the three-dimensional or projected galaxy density (a sparse, biased tracer) or the convergence field (a weighted, projected version of it). In this section we describe our modelling of the shear and galaxy correlation signal of troughs by the interrelation of these observables and the underlying matter density.

We assume that the three-dimensional galaxy field can be described as a deterministic, biased tracer of the matter. This means that the $3$D contrast $\delta$ of matter density $\rho$,
\begin{equation}
\delta=\frac{\rho-\langle\rho\rangle}{\langle\rho\rangle} \; ,
\end{equation}
and the equivalent quantity defined for the galaxy field are proportional at any position. Their ratio defines the bias $b$, which depends on the galaxy population.

Lensing convergence $\kappa$ is related to $\delta$ via the projection integral over comoving distance $\chi$ \citep[cf., e.g.][]{2001PhR...340..291B},
\begin{equation}
\kappa(\bm{\theta}) = \int_0^\infty \mathrm d \chi\ q_{\kappa}(\chi) \ \delta(\chi\bm{\theta},  \chi)\ ,
\label{eqn:kappadef}
\end{equation}
where
\begin{equation}
q_\kappa(\chi) = \frac{3H_0^2\Omega_{\rm m}^0}{2} \, \chi \, \mathcal{G}(\chi)
\end{equation}
with
\begin{equation}
\mathcal{G}(\chi) = \int_\chi^{\infty} \mathrm{d}\chi'\ n_{\rm source}(\chi') \frac{\chi'-\chi}{\chi'}\ .
\end{equation}
Here, $n_{\rm source}(\chi)$ is the distribution of source galaxies of the lensing measurement.

An overdensity of convergence inside a circular aperture relative to its edge results in a tangential alignment of background galaxy shapes. Correspondingly, an underdensity causes radial alignment. Both cases are described by \citep[cf.][p.~279f]{saasfee}
\begin{equation}
\gamma_{\rm t}(\theta)=\langle{\kappa}\rangle(<\theta)-\kappa(\theta) \; .
\label{eqn:gammakappa}
\end{equation}
Here, $\gamma_{\rm t}(\theta)$ is the tangential component of gravitational shear averaged over the edge of a circle of radius $\theta$, $\kappa(\theta)$ is the equivalent average of the convergence and $\langle{\kappa}\rangle(<\theta)$ is the mean convergence inside the circle. For the case of $|\gamma_{\rm t}|\ll1$, $|\kappa|\ll1$, tangential components of gravitational shear and reduced shear $g_{\rm t}$ are approximately equal and observable as the mean tangential alignment of background galaxy ellipticity. The cross component of shear, $\gamma_{\times}$, rotated by $45^{\circ}$ relative to the tangential direction, is expected to be zero when taking the average over the full circle for a single thin lens or over an ensemble of thick lenses. 

In order to connect these fields and model the trough signal, we make these three assumptions:
\begin{itemize}
\item We apply the \citet{1954ApJ...119..655L} approximation to compute the angular power spectrum of the \emph{projected matter density contrast} $\delta_{\Sigma}(\bm{\theta})$ within the redshift range of the redMaGiC galaxies used for the trough selection ($0.2 \leq z \leq 0.5$). The same approximation is also used to compute the cross power spectrum between $\delta_{\Sigma}$ and the convergence field $\kappa(\bm{\theta})$ relative to the background galaxy redshift distribution.
\item We assume that $\delta_{\Sigma}(\bm{\theta})$ and $\kappa(\bm{\theta})$ follow a Gaussian distribution -- at least when they are averaged over the trough radius or over the annuli in which we measure the shear.
\item We assume that the redMaGiC galaxies are placed on to $\delta_{\Sigma}$ via a biased Poisson process.
\end{itemize}
In this section, we describe how these assumptions translate to a prediction for the expected shear signal and galaxy density around troughs.

\subsection{Projected matter density and galaxy counts}

Let the volume density of redMaGiC galaxies as a function of comoving distance $\chi$ be given by $n_{\rm lens}(\chi)$. The projected galaxy contrast is proportional to a weighted projection of matter contrast $\delta_{\Sigma}$. In a flat universe, the latter is calculated as
\begin{equation}
\delta_{\Sigma}(\bm{\theta}) = \frac{\Sigma-\bar\Sigma}{\bar\Sigma} = \int_{\chi_0}^{\chi_1} \mathrm d \chi\ n_{\rm lens}(\chi) \ \delta(\chi\bm{\theta}, \chi)\ , 
\end{equation}
where $\delta(\chi\bm{\theta}, \chi)$ is the 3D matter contrast at the point $(\chi\bm{\theta}, \chi)$ on the backward light cone. For a galaxy sample with constant comoving density, such as the redMaGiC catalogue, this is a simple volume weighting of matter density, $n_{\rm lens}(\chi)\propto \mathrm{d}V/[\mathrm{d}\Omega\, \mathrm{d}\chi]$, and $\delta_{\Sigma}$ is the common projected matter density contrast.

If we average $\delta_{\Sigma}$ over circles of angular radius $\theta_{\rm T}$, we arrive at the new random field $\delta_{\rm T}$,
\begin{equation}
\delta_{\rm T}(\bm{\theta}) = \frac{1}{\uppi\theta_{\rm T}^2} \int_{|\bm{\theta} - \bm{\theta}'|\leq \theta_{\rm T}} \mathrm{d}^2 \bm{\theta}'\ \delta_{\Sigma}(\bm{\theta}') \; .
\end{equation}
In this we have made the approximation of a flat sky, valid for $\theta_T\ll1$.

If the galaxies are placed on to $\delta_{\Sigma}(\bm{\theta})$  via a biased Poisson process, then the discrete probability $P$ of finding $N$ galaxies within $\theta_{\rm T}$ given the value of $\delta_{\rm T}$ is
\begin{equation}
P(N|\delta_{\rm T}) = \frac{1}{N!} \left(\bar N \left[1 + b\delta_{\rm T} \right] \right)^{N} \, \exp\left(-\bar N \left[1 + b\delta_{\rm T} \right] \right)\ .
\label{eqn:pN}
\end{equation}
For $\delta_{\rm T} < -\frac{1}{b}$ we assume $P(N>0|\delta_{\rm T}) = 0$ and $P(N=0|\delta_{\rm T}) = 1$ (see also Appendix A). We have used the bias $b$ and mean galaxy count $\bar N$ within $\theta_{\rm T}$. For our model predictions shown later, we fix the bias at a fiducial value of $b=1.6$ or vary it between $1.4,\ldots,1.8$ to show the dependence on bias in the relevant range. Note that we neglect the moderate redshift dependence of the bias of redMaGiC galaxies \citep[cf.][]{redmagic}, which is a good approximation for the limited redshift range used here. 

We identify troughs as circles in the sky with low galaxy count $N$. Given any $N$, the expected value of $\delta_{\rm T}$ is
\begin{equation}
\langle \delta_{\rm T} | N \rangle = \int_{-1}^\infty \mathrm{d} \delta_{\rm T}\ \delta_{\rm T}\ p(\delta_{\rm T} | N) \; .
\end{equation}
Bayes' theorem tells us that
\begin{equation}
\label{eq:Bayes}
p(\delta_{\rm T} | N) = \frac{P(N|\delta_{\rm T})\, p(\delta_{\rm T})}{P(N)} \ .
\end{equation}
When the variance $\Var(\delta_{\rm T}):=\sigma^2_{\rm T} \ll 1$, we can approximate $p(\delta_{\rm T})$ by a Gaussian distribution, i.e.
\begin{equation}
p(\delta_{\rm T}) = \frac{1}{\sqrt{2\uppi\sigma^2_{\rm T}}}\exp\left(-\frac{\delta_{\rm T}^2}{2\sigma^2_{\rm T}}\right)\ .
\label{eqn:pdeltat}
\end{equation}

In Appendix B, we derive how $\sigma_{\rm T}^2$ can be calculated from the power spectrum of $\delta_{\Sigma}$, which in the \citet{1954ApJ...119..655L} approximation is related to the $3$D matter power spectrum $P_\delta$ by \citep[cf., e.g.][Eqn.~2.84]{2001PhR...340..291B}
\begin{equation}
C_\Sigma(\ell) = \int_0^\infty \mathrm d \chi\ \frac{n_{\rm lens}(\chi)^2}{\chi^2} \ P_\delta\left(\frac{\ell}{\chi},  \chi\right)\ .
\end{equation}
For our calculations, we use the non-linear matter power spectrum $P_\delta$ from \citet{2003MNRAS.341.1311S}.

\begin{figure}
\centering
\includegraphics[width=0.48\textwidth]{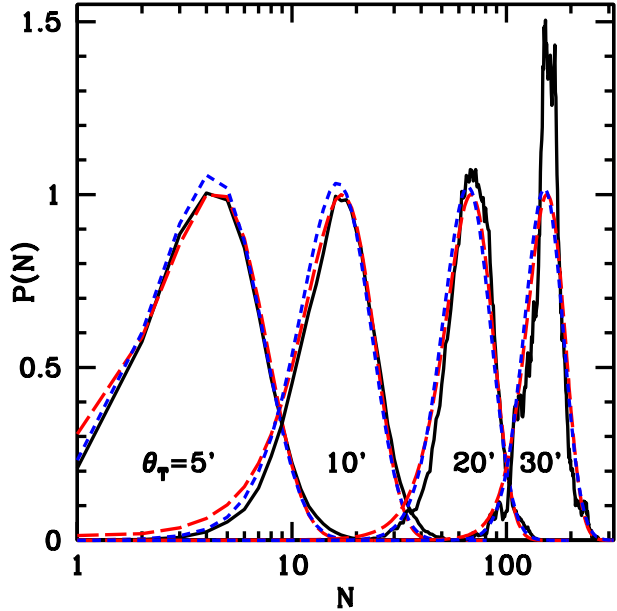}
\caption{Distribution of redMaGiC galaxy counts in $z_{\rm T}\in[0.2,0.5]$ for various trough radii $\theta_{\rm T}$. We show measurements (black, solid line) and model predictions for Gaussian matter fields on to which galaxies are placed as biased tracers with $b=1.6$ and Poissonian noise (red, long-dashed lines). Model predictions for a lognormal matter density are also shown (blue, short-dashed lines). Normalization of $P(N)$ is matched to make ${\rm max}(P^{\rm Gaussian})=1$ for each trough radius.}
\label{fig:pn}
\end{figure}

The normalization factor in equation \ref{eq:Bayes}, $P(N)$, gives the probability of finding $N$ galaxies inside a cylinder of radius $\theta_{\rm T}$. Fig.~\ref{fig:pn} shows the observed distribution of galaxy counts in cylinders. There is reasonable agreement to our prediction from a Gaussian random field $\delta_{\rm T}$ on to which galaxies are placed by means of equation~\ref{eqn:pN} (dashed, red lines). The distribution of counts inside 30~arcmin radii appears more peaked than the model, potentially explained by the fact that there is too little area to have enough uncorrelated troughs at this scale.

As an alternative prescription, we also calculate the expected $P(N)$ for the case of a lognormally distributed matter density. This is done by replacing the Gaussian $p(\delta_{\rm T})$ of equation~\ref{eqn:pdeltat} by a shifted lognormal distribution of the same variance with a minimum value of $\delta_{\rm T}=-1$. The result (dotted, blue lines in Fig.~\ref{fig:pn}) resembles the low-density tail of the galaxy count distribution more closely. 

\subsection{Convergence and shear profile}

Our goal is to compute the mean value of $\kappa$ at a distance $\theta$ from the trough centre. For this, we define a set of annuli $i=1,\ldots,n$ around the trough for which $\theta\in[\theta_i,\theta_{i+1})$. Let $K_i$ be the average of $\kappa$ in annulus $A_i$, 
\begin{equation}
K_i(\boldsymbol \theta) = \frac{1}{\uppi(\theta_{i+1}^2-\theta_i^2)} \int_{A_i} \mathrm d^2 \bm{\theta}'\ \kappa(\boldsymbol \theta') \; .
\label{eqn:K}
\end{equation}
If both $\delta_{\rm T}$ and $K_i$ have Gaussian distributions with zero mean, then the expectation value of $K_i$ for a fixed value of $\delta_{\rm T}$ is given by
\begin{equation}
\langle K_i | \delta_{\rm T} = s \rangle = \frac{\Cov(\delta_{\rm T}, K_i)}{\sigma^2_{\rm T}}\, s\; ,
\label{eqn:Ks}
\end{equation}
where the covariance $\Cov(\delta_{\rm T}, K_i)$ can be computed in terms of the cross power spectrum $C_{\kappa, \Sigma}(\ell)$ of $\Sigma$ and $\kappa$ (see Appendix B). Note that the Gaussian approximation for the matter contrast and convergence becomes accurate regardless of the pointwise $p(\delta)$ since all random fields are smoothed over annuli or circles and a large redshift range.

The expectation value of $K_i$ when the trough contains $N$ galaxies is finally given by
\begin{align}
\langle K_i | N \rangle &= \int_{-1}^{\infty} \mathrm{d}s\ \langle K_i | \delta_{\rm T} = s \rangle\ p(\delta_{\rm T} = s|N) \nonumber \\
&= \frac{\Cov(\delta_{\rm T}, K_i)}{\sigma^2_{\rm T}} \int_{-1}^{\infty} \mathrm{d}s\ s\ p(\delta_{\rm T} = s|N) \nonumber \\
&= \frac{\Cov(\delta_{\rm T}, K_i)}{\sigma^2_{\rm T}}\, \langle \delta_{\rm T} | N \rangle\ .
\end{align}

If we select as troughs all cylinders with $N\leq N_{\rm max}$ galaxies, the mean $K_i$ around them will be
\begin{equation}
\label{eq:kappa_in_annulus}
\langle K_i | \leq N_{\rm max} \rangle = \frac{\Cov(\delta_{\rm T}, K_i)}{\sigma^2_{\rm T}}\, \frac{\sum_{N = 0}^{N_{\rm max}} P(N) \; \langle \delta_{\rm T} | N \rangle}{\sum_{N=0}^{N_{\rm max}} P(N) }\; .
\end{equation}
The tangential shear signal around troughs is given by equation~\ref{eqn:gammakappa}, using equation~\ref{eq:kappa_in_annulus} to calculate the mean convergence in each annulus.

\subsection{Trough-galaxy angular correlation}
\label{sec:modelw}

The trough-galaxy angular correlation function can be modelled in a very similar way. First define annuli $A_i$ around the trough that correspond to the bins in which $w(\theta)$ is measured. The mean density contrast $w_i$ in each annulus is given by (cf.~equation~\ref{eqn:K})
\begin{equation}
w_i(\boldsymbol \theta) = \frac{1}{\uppi(\theta_{i+1}^2-\theta_i^2)} \int_{A_i} \mathrm d^2 \bm{\theta}'\ \delta_\Sigma(\boldsymbol \theta')\; .
\end{equation}
Under the assumptions of Gaussianity, the expectation value of $\delta_i$ for a fixed value of $\delta_{\rm T}$ is given by (cf.~equation~\ref{eqn:Ks})
\begin{equation}
\langle w_i | \delta_{\rm T} = s \rangle = \frac{\Cov(\delta_{\rm T}, w_i)}{\sigma^2_{\rm T}}\, s \; .
\end{equation}
In analogy to equation \ref{eq:kappa_in_annulus}, the mean density contrast in annulus $A_i$ around the trough is given by 
\begin{equation}
\label{eq:delta_in_annulus}
\langle w_i | \leq N_{\rm max} \rangle = \frac{\Cov(\delta_{\rm T}, w_i)}{\sigma^2_{\rm T}}\, \frac{\sum_{N = 0}^{N_{\rm max}} P(N) \; \langle \delta_{\rm T} | N \rangle}{\sum_{N=0}^{N_{\rm max}} P(N) }\; .
\end{equation}

The average number of galaxies in an annulus $i$ \emph{outside} the trough radius is given by
\begin{align}
\label{eq:count_profile}
\langle N_{i} | \leq N_{\rm max} \rangle_{\mathrm{out}} &= \bar N \frac{A_i}{A_{\rm T}} [1 + b\langle w_i | \leq N_{\rm max} \rangle] \nonumber \\
&= \bar N_i [1 + b\langle w_i | \leq N_{\rm max} \rangle]\ ,
\end{align}
where the mean galaxy count $\bar N_i$ in annulus $i$ is obtained by rescaling the average galaxy count inside one trough radius to the area $A_i$ of the annulus,
\begin{equation}
\bar N_i = \bar N \frac{A_i}{A_{\rm T}}\ .
\end{equation}
The profile of galaxy counts around a trough is then given by
\begin{align}
\langle w_{N, i} | \leq N_{\rm max} \rangle_{\mathrm{out}} &= \frac{\langle N_i | \leq N_{\rm max} \rangle_{\mathrm{out}}}{\bar N_i} - 1 \nonumber \\
&= b \langle w_{i} | \leq N_{\rm max} \rangle\ .
\end{align}

The situation is more complicated for annuli \emph{inside} the trough radius $\theta_{\rm T}$. Here, the Poisson noise of the different bins is correlated. This is because the sum of the galaxy counts in the different bins has to meet the requirement by which we selected the troughs.

Without full treatment of the covariances, we make a prediction for the galaxy number counts inside the trough that (i) matches the mean galaxy counts predicted by the full model and (ii) matches the projected matter contrast profile with a given bias. To this end, we simply replace $\bar N$ in equation \ref{eq:count_profile} by the predicted mean number of galaxies inside \emph{selected} troughs, e.g. when demanding that $N_{\rm T} \leq N_{\rm max}$. This number is given by
\begin{equation}
\bar N_{\rm T} = \frac{\sum_{N = 0}^{N_{\rm max}} P(N) \; N}{\sum_{N=0}^{N_{\rm max}} P(N) }\ .
\end{equation}
The mean number of galaxies found in an annulus inside the trough is then given by
\begin{equation}
\langle N_{i} | \leq N_{\rm max} \rangle_{\mathrm{in}} = \bar N_{\rm T} \frac{A_i}{A_{\rm T}} [1 + b\langle w_i | \leq N_{\rm max} \rangle]
\end{equation}
and the profile of galaxy counts inside the trough radius is given by
\begin{align}
\langle w_{N, i} | \leq N_{\rm max} \rangle_{\mathrm{in}} &= \frac{\langle N_i | \leq N_{\rm max} \rangle_{\mathrm{in}}}{\bar N_i} - 1 \nonumber \\
&= \frac{\bar N_T}{\bar N}[ 1 + b \langle w_{i} | \leq N_{\rm max} \rangle ] - 1\ .
\end{align}

\section{Measurement}

In the following section, we correlate trough positions with the shear signal of background galaxies (Section~\ref{sec:g}). Additionally, we measure the projected number density profile of redMaGiC galaxies in the same redshift and luminosity range used for the trough selection, i.e.~the angular two-point cross-correlation of troughs and galaxies (Section~\ref{sec:w}). 

\subsection{Shear signal}
\label{sec:g}

\begin{figure*}
\centering
\includegraphics[width=0.48\textwidth]{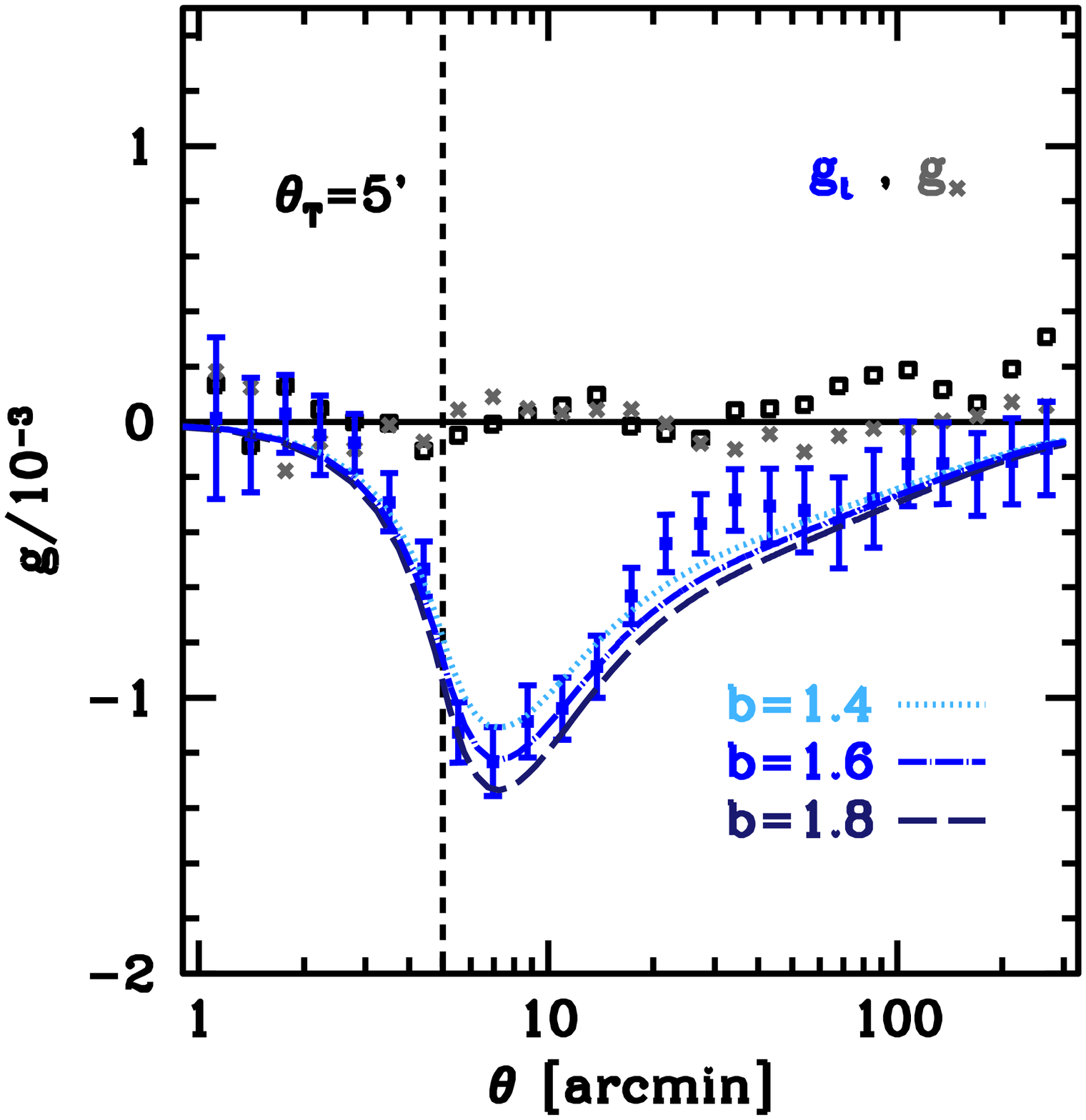}
\includegraphics[width=0.48\textwidth]{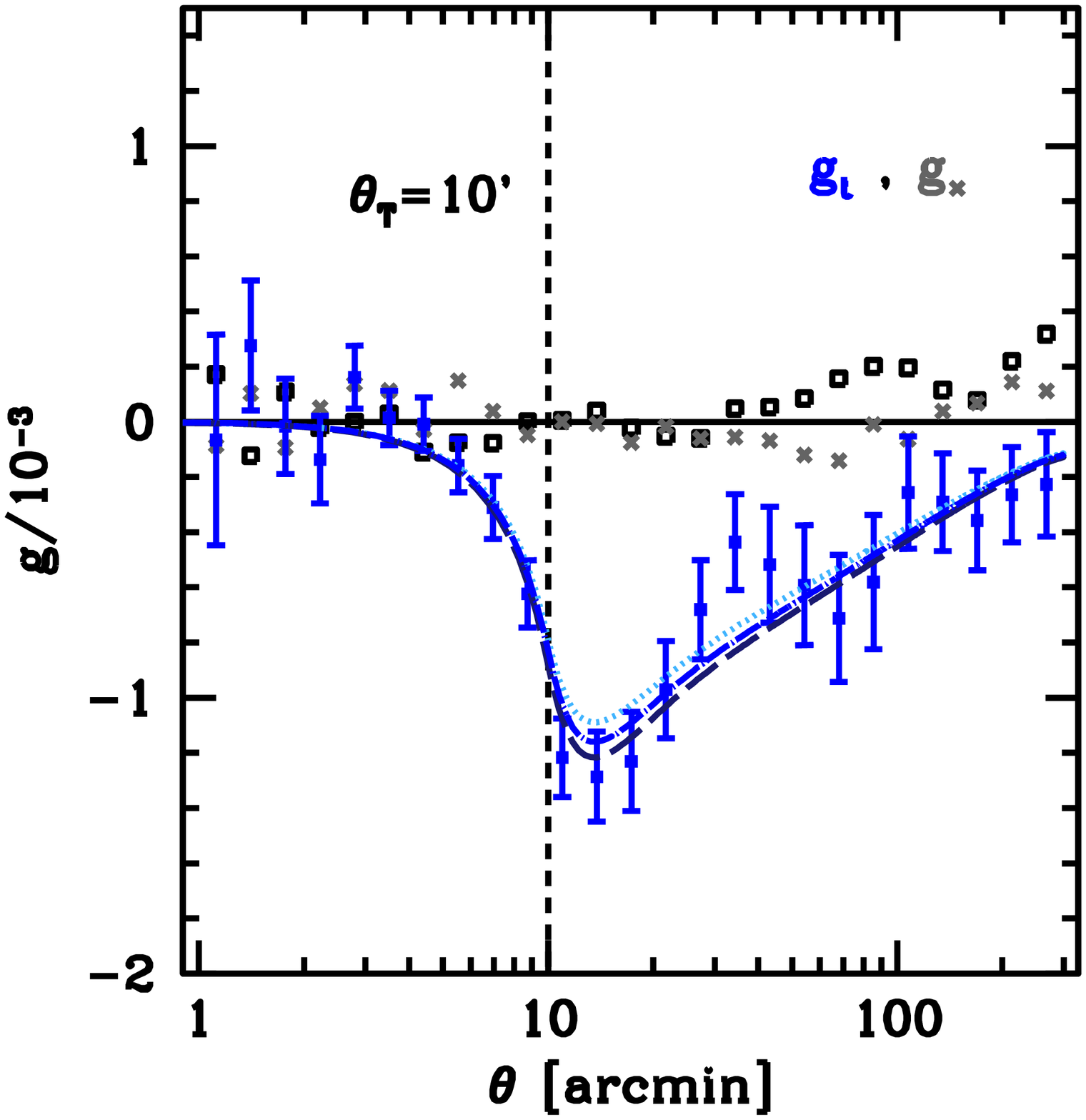}
\includegraphics[width=0.48\textwidth]{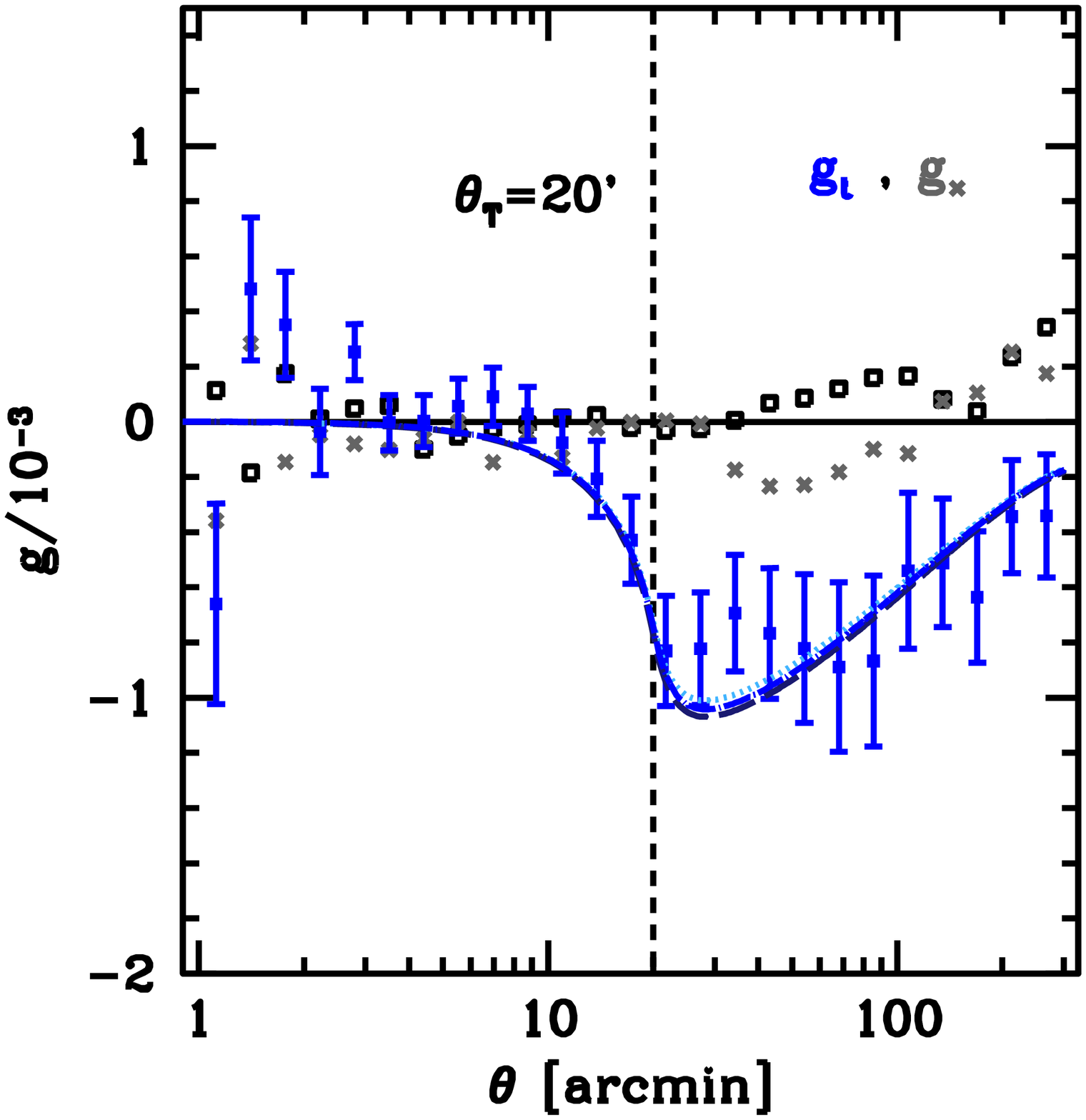}
\includegraphics[width=0.48\textwidth]{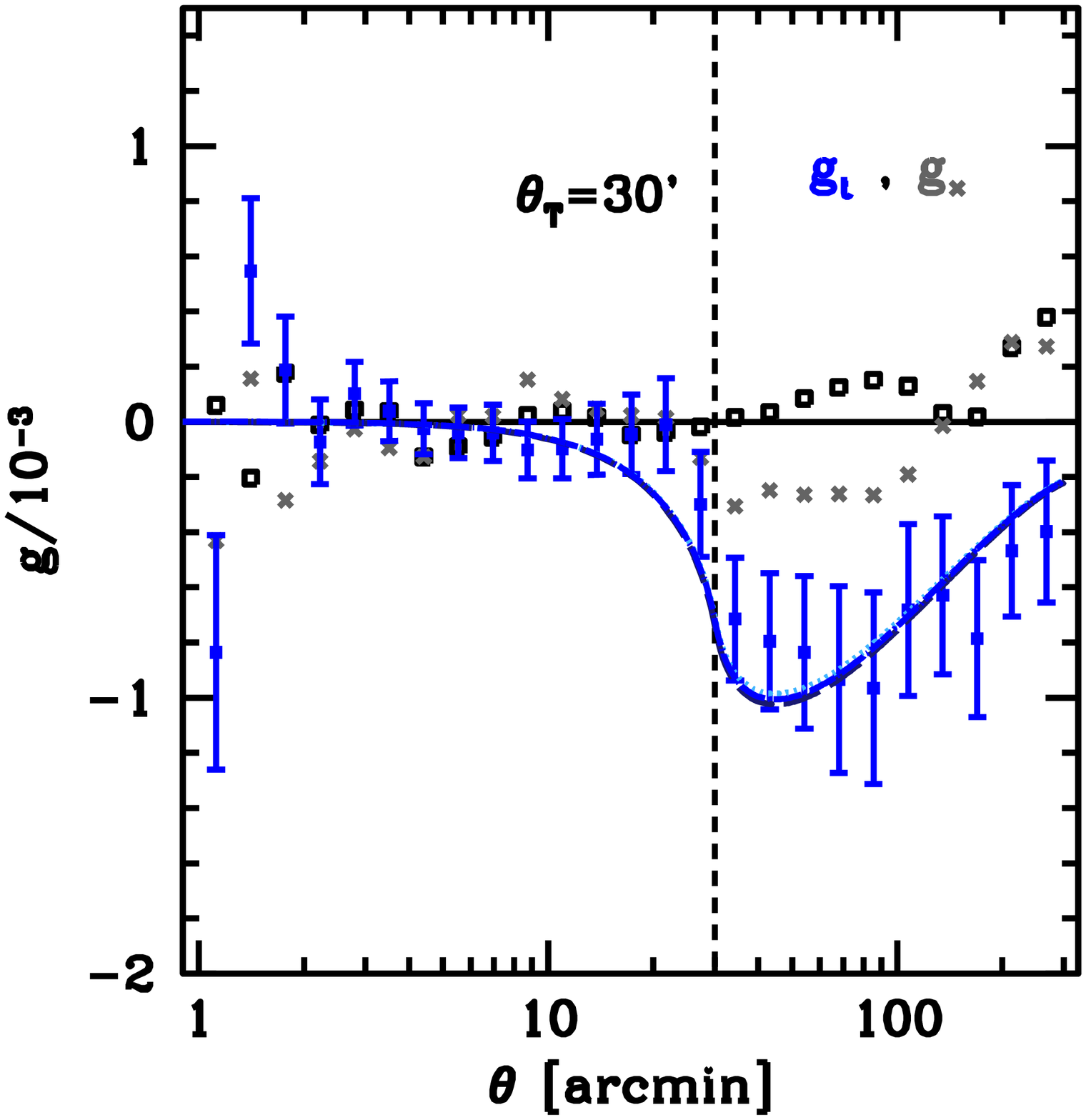}
\caption{Weak lensing signal of galaxy troughs of $\theta_{\rm T}=$5, 10, 20 and 30~arcmin radius (top left to bottom right). Shown is the tangential shear signal (blue) around points from the lower 20th percentile in galaxy counts in cylinders of $z=0.2,\ldots,0.5$. Lines show model predictions (cf.~Section~\ref{sec:model}) for our fiducial cosmology and, for illustration of the bias dependence, a bias of $b=1.4,1.6,1.8$ (light to dark blue, dotted, dot-dashed and dashed lines). Cross-shear is shown with grey cross symbols, to be interpreted with error bars of similar size. Tangential shear around random points, subtracted from the trough measurement, is shown with black open symbols.}
\label{fig:g}
\end{figure*}

We measure the mean shear of background galaxies around troughs, selected as described in Section~\ref{sec:troughselection}. To correct for potential additive shear systematic errors, we subtract the tangential shear measured around random points. Since the masked region depends on the respective trough radius, the random shears for each $\theta_{\rm T}$ differ slightly. Fig.~\ref{fig:g} shows measured tangential and cross shears. 

Per-mille radial alignment of background galaxies at and beyond the trough radius is detected with high significance in all bins (cf. Section~\ref{sec:significance}, Table~1). Cross shears are consistent to the expected null signal within the uncertainties (cf. also the reduced $\chi^2_{\times}$ in Table~1). The model proposed in Section~\ref{sec:model} is a good fit to the data in all bins (cf. Section~\ref{sec:modelfit} and reduced $\chi^2_{\rm mod}$ in Table~1).

\subsubsection{Tomography}
\label{sec:tomo}
By splitting either the source sample or using smaller redshift ranges for selecting the troughs, it is possible to probe the redshift evolution of the trough lensing signal. We perform both measurements in the following.

\begin{figure*}
\centering
\includegraphics[width=0.48\textwidth]{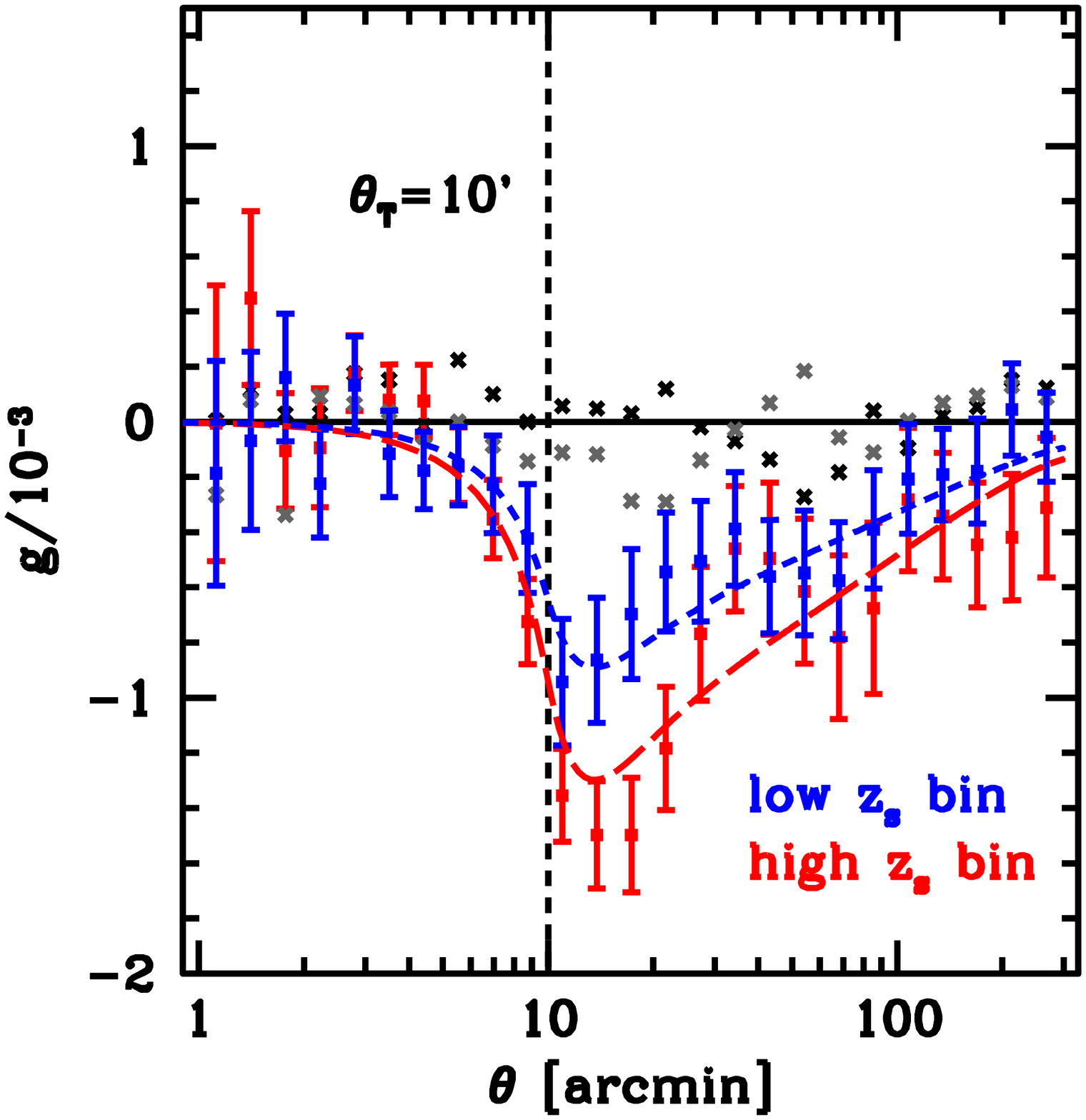}
\includegraphics[width=0.48\textwidth]{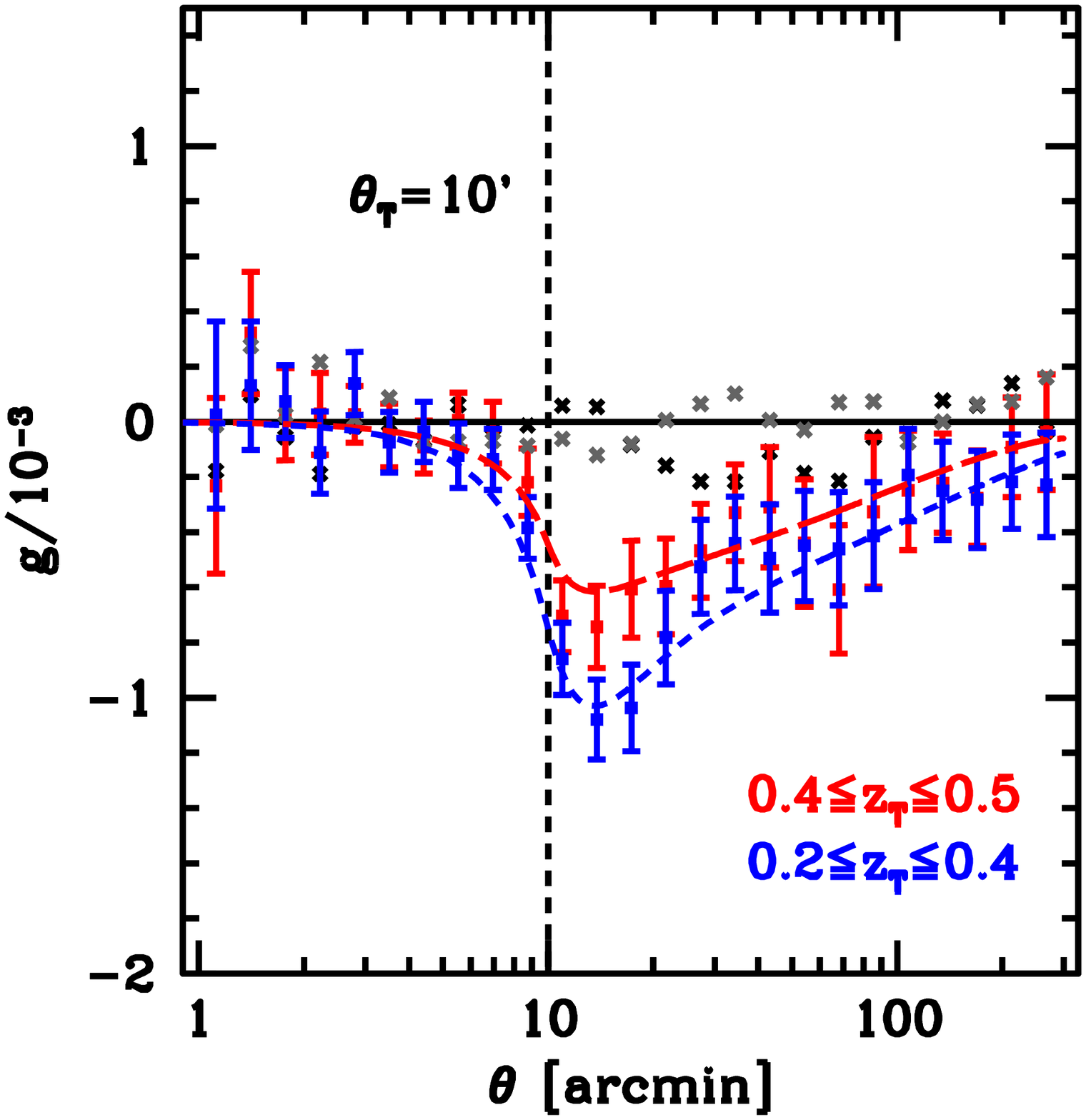}
\caption{Source tomography (left) and trough redshift range tomography (right) of trough lensing signal. Left-hand panel shows tangential shear signal for sources in the lower (blue, short dashed) and higher (red, long dashed) of our redshift bins (cf.~Section~\ref{sec:sourcecat}). Right-hand panel shows signal of troughs selected by the galaxy count in $z_T=0.2,\ldots,0.4$ (blue, short dashed) and $z=0.4,\ldots,0.5$ (red, long dashed).  Model predictions (cf.~Section~\ref{sec:model}) are shown for a bias of $b=1.6$ and grey/black points indicate $g_{\times}$ for both measurements.}
\label{fig:gtomo}
\end{figure*}
\begin{figure*}
\centering
\includegraphics[width=0.48\textwidth]{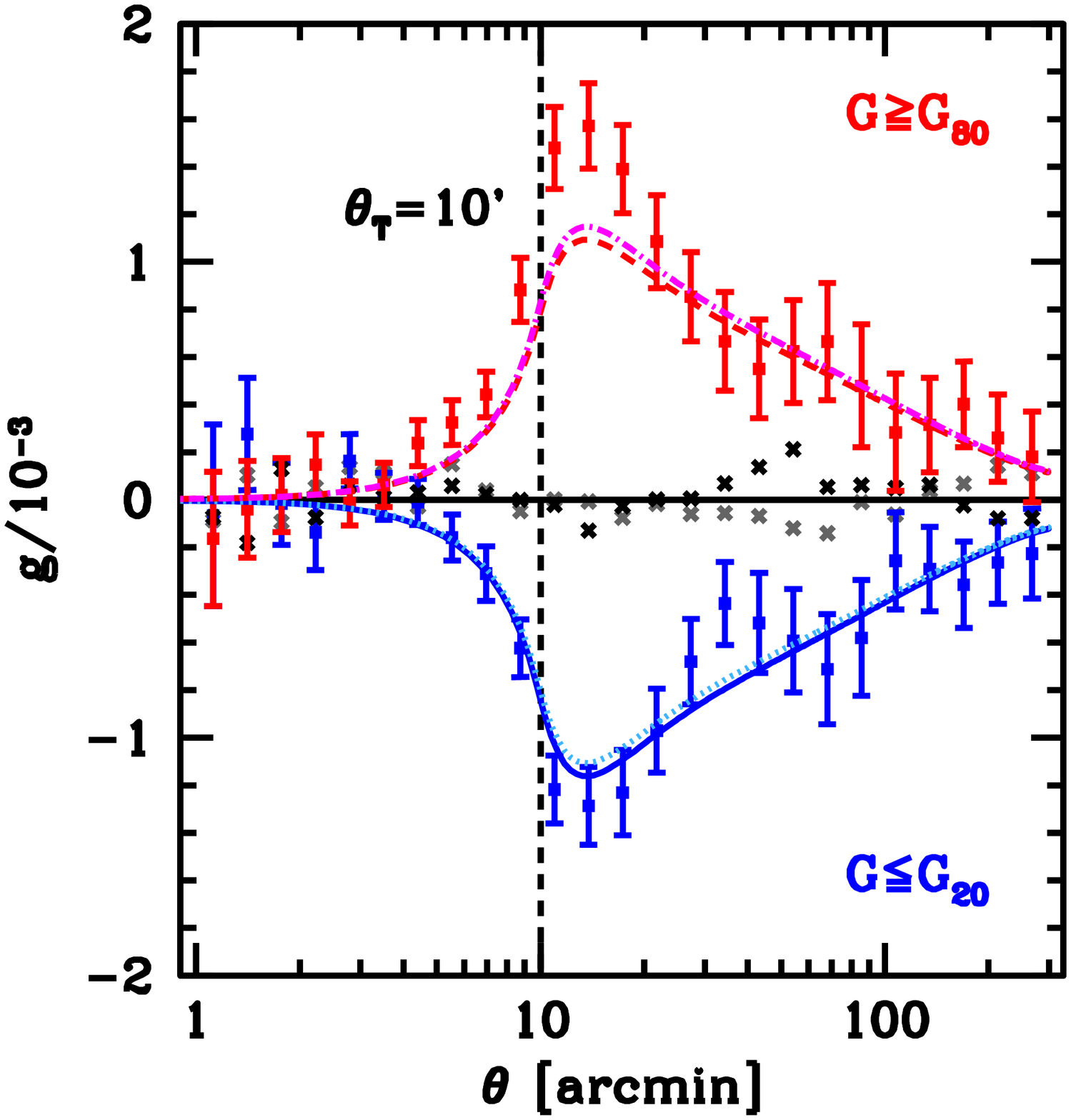}
\includegraphics[width=0.48\textwidth]{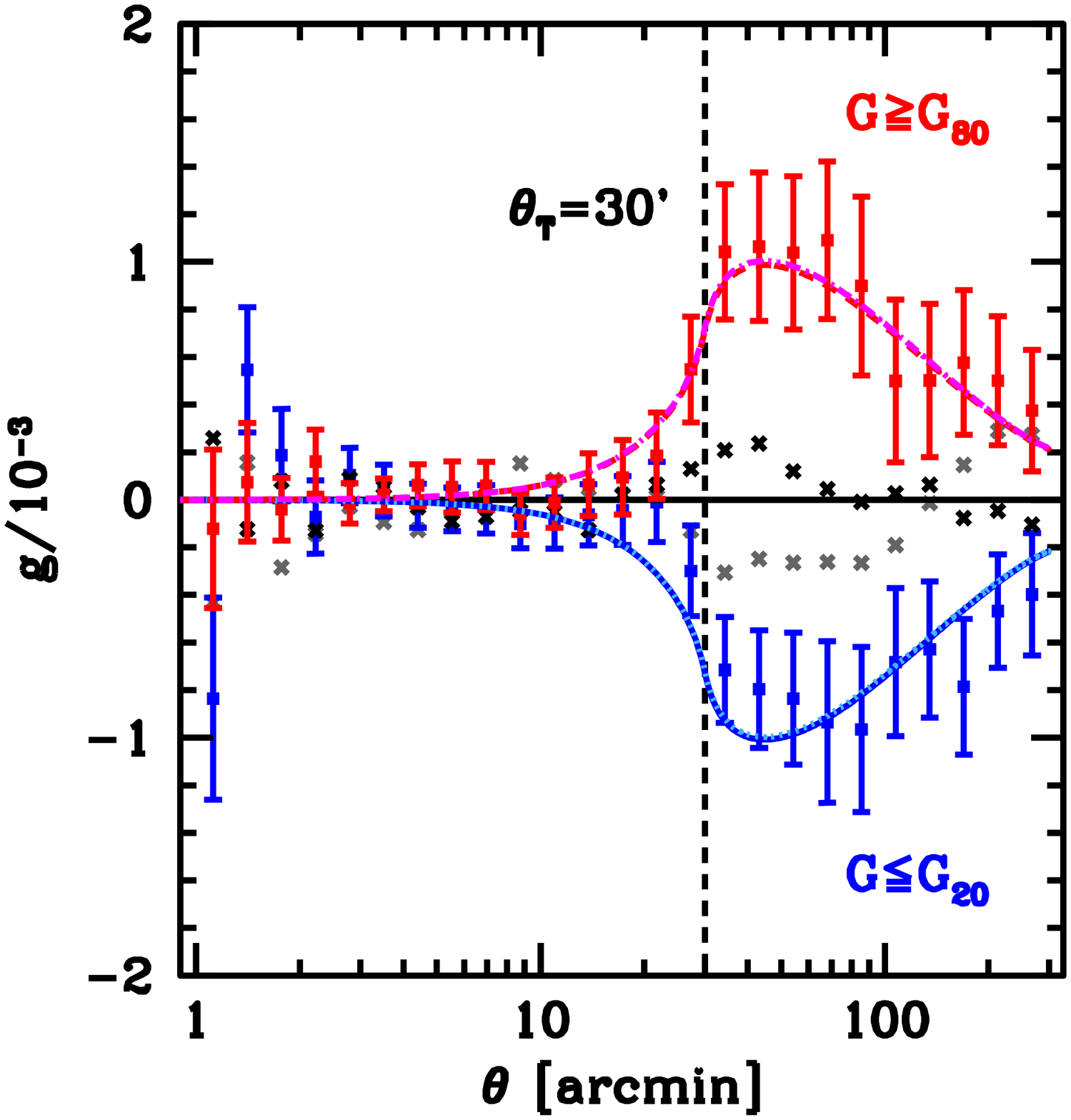}
\caption{Tangential shear signal for troughs, i.e.~centres of cylinders below the 20th (measurement and model in blue), and overdense cylinders above the 80th percentile (red) in galaxy count. We plot model predictions for a bias of $b=1.6$, with solid (dashed) lines assuming a Gaussian and dotted (dashed-dotted) lines a lognormal distribution of the matter contrast around troughs (overdense cylinders). Grey/black points indicate $g_{\times}$ for both measurements}
\label{fig:gperc}
\end{figure*}

\begin{itemize}
\item For source tomography, we divide the source galaxy sample into two redshift bins (cf.~Section~\ref{sec:sourcecat}). Note that since troughs are thick lenses, the change in source redshift causes more than a simple change in amplitude. The differential weighting as a function of lens redshift inside the $z=0.2\ldots0.5$ cylinder also influences the shape of the shear profile. Due to the nearly power-law matter two-point correlation at all redshifts, however, the latter effect is small. The left-hand panel of Fig.~\ref{fig:gtomo} shows the source-tomographic signal. We note that the agreement of the measurement with the model in both bins is additional evidence for the appropriateness of the $p(z)$s as estimated for our source samples \citep[cf.][]{photoz}.
\item For trough redshift tomography, we split the trough redshift range into two approximately equal-volume slices $z=0.2\ldots0.4$ and $z=0.4\ldots0.5$. When using these smaller redshift ranges for the trough selection, two effects reduce the SNR: (1) due to the lower galaxy count, Poissonian noise weakens the correlation of trough positions with matter underdensity; and (2) uncorrelated, overdense large-scale structure along the line-of-sight outside the trough redshift range causes additional variance in the lensing signal. Shear measurements are shown in the right-hand panel of Fig.~\ref{fig:gtomo}. The signal is reduced as expected, but the measurement is still highly significant and consistent with the model in both cases (see Table~1 for details on significance and goodness of fit).
\end{itemize}

\subsubsection{Galaxy density percentiles}
\label{sec:symmetry}
All measurements presented above use troughs selected to be below the lower 20th percentile of galaxy counts. Measurements with larger limiting percentiles (e.g.~the 30th percentile) give results of similar significance but smaller amplitude. 

It is particularly interesting, however, to study the symmetry of matter in the overdense and underdense tails of the galaxy field. For dense enough tracers and large enough scales, the expectation is that all involved fields are approximated well by a Gaussian distribution. This should lead to symmetric shear signals at the same upper and lower percentiles. On smaller scales, the galaxy counts (if only due to Poisson noise) and the matter density and convergence field (since $|\delta|\ll1$ is no longer true and non-linear evolution boosts high-density fluctuations) deviate from a Gaussian distribution and we expect some degree of asymmetry between the low- and high-density signal.

The measurement for both the lower and upper 20th percentile is shown in Fig.~\ref{fig:gperc} and is in agreement with these expectations. At small trough radii, there appears to be a significant asymmetry, with the overdense regions showing a larger shear signal than anticipated from our model or the measurement of underdensities. For larger cylinders, such an effect is not detected. A lognormal model of the matter contrast (dashed lines) makes virtually no difference for larger trough radii. For smaller trough radii, the shears around high-density cylinders are predicted to be somewhat larger, yet not sufficiently so to fit the data well. We hypothesize that the discrepancy between high- and low-density cylinders can rather be explained by an environment dependence of the bias of the redMaGiC tracer galaxies: because the mean bias of galaxies in overdense regions is larger than in underdense regions, the shear around small, high-density cylinders gets boosted relative to the signal around the low-density troughs (cf.~the bias dependence of the model prediction in Fig.~\ref{fig:g}).

\subsubsection{Significance}
\label{sec:significance}

For estimating uncertainties, we use a set of $N_{\rm j}=100$ jackknife resamplings. In order to ensure that these are approximately equally populated with troughs, we choose them with a K-means algorithm\footnote{\texttt{https://github.com/esheldon/kmeans\_radec/}} on the catalogue of 5~arcmin trough positions. The delete-one jackknife yields a covariance
\begin{equation}
\Cov(f_1,f_2)=\frac{N_{\rm j}-1}{N_{\rm j}}\sum_{i=1}^{N_{\rm j}}(f_{1,\neg i}-\langle f_1 \rangle)(f_{2,\neg i}-\langle f_2 \rangle)
\label{eqn:cov} 
\end{equation}
for two quantities $f_1,f_2$ estimated from the data excluding region $i$ ($f_{\neg i}$) or averaging over all ($\langle f\rangle=N^{-1}\sum_{i=1}^N f_{\neg i}$). In our case, we estimate the covariance matrix $\mathsf{\mathbf{C}}$ of tangential shear measurements (or, in Section~\ref{sec:w}, angular two-point correlation measurements) in our set of angular bins. 

\begin{figure}
\centering
\includegraphics[width=0.48\textwidth]{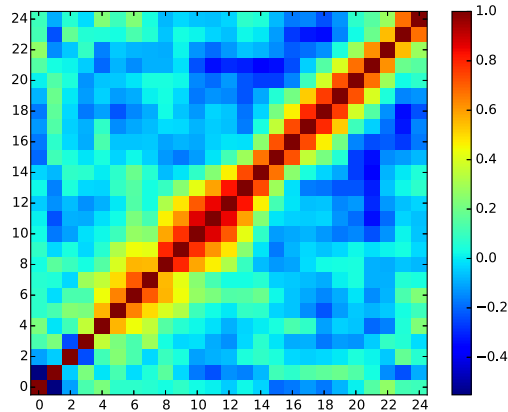}
\caption{Correlation matrix $R_{ij}$ of shear around 10~arcmin troughs measured in the logarithmic angular bins of Fig.~\ref{fig:g} as estimated from 100 jackknife regions.}
\label{fig:cov}
\end{figure}

Fig.~\ref{fig:cov} shows the correlation coefficients $R_{ij}=\Cov(g_{\rm t}^i,g_{\rm t}^j)/\sqrt{\Var(g_{\rm t}^i)\Var(g_{\rm t}^j)}$ estimated for our fiducial 10~arcmin trough measurement. At intermediate and large radii, neighbouring bins are highly positively correlated, which is even more the case for the larger troughs. The negative correlation of the innermost bins is a generic feature that appears in all trough sizes probed and is connected to the opposite sign of the first two data points of the lower panels of Fig.~\ref{fig:g}. Both this and the off-diagonal negative correlations at large radii are also seen in less noisy versions of the covariance determined from simulations (cf.~Friedrich et al., in preparation).

\begin{table}
\begin{center}
\begin{tabular}{|l|l|l|l|r|r|r|r|}
\hline
\multicolumn{3}{|c|}{Trough selection} & & \multicolumn{2}{|c|}{Significance} & \multicolumn{2}{|c|}{Reduced $\chi^2$} \\
$\theta_{\rm T}$ & $z_{\rm T}\in$ & $\mathcal{P}$  & $z_s$ & $g_{\rm t}(\theta_{\rm T})$ & $\Gamma$ & $\chi^2_{\rm mod}$ & $\chi^2_{\times}$ \\ \hline \hline
 5 & $[0.2,0.5]$ & $\leq0.2$  & all & 10                     & 17                          &               1.1    &         0.3       \\ \hline
10 & $[0.2,0.5]$ & $\leq0.2$  & all &  9                     & 12                          &               1.4    &         0.5       \\ \hline
20 & $[0.2,0.5]$ & $\leq0.2$  & all &  4                     &  9                          &               0.9    &         1.0       \\ \hline
30 & $[0.2,0.5]$ & $\leq0.2$  & all &  3                     &  6                          &               0.7    &         1.0       \\ \hline \hline
10 & $[0.2,0.5]$ & $\leq0.2$  & low &  4                     &  6                          &               0.6    &         0.5       \\ \hline
10 & $[0.2,0.5]$ & $\leq0.2$  & high & 8                     & 11                          &               1.4    &         0.6       \\ \hline \hline
10 & $[0.2,0.4]$ & $\leq0.2$  & all &  7                     &  9                          &               0.9    &         0.4       \\ \hline
10 & $[0.4,0.5]$ & $\leq0.2$  & all &  5                     & 10                          &               1.2    &         0.6       \\ \hline \hline
10 & $[0.2,0.5]$ & $\geq0.8$  & all &  9                      & 12                         &               1.0    &         0.4       \\ \hline
\end{tabular}
\end{center}
\caption{Metrics of significance of detection of shear around troughs of radius $\theta_{\rm T}$ selected from the galaxy field in the given redshift range $z_{\rm T}$ at the percentile threshold $\mathcal{P}$ for sources in the indicated $z_{\rm s}$ bins. We list the SNR of shear at the trough radius, $g_{\rm t}(\theta_{\rm T})/\sigma_{g_{\rm t}(\theta_{\rm T})}$, and of the optimally weighted linear combination of shears, $\Gamma/\sigma_{\Gamma}$. See description in Section~\ref{sec:significance} for details. The remaining columns show the reduced $\chi^2$ of the residuals of model and measurement (cf. Section~\ref{sec:modelfit}) and of cross-shears.}
\label{tbl:significance}
\end{table}

We ensure the significances defined below are stable under a change of binning scheme and jackknife regions by calculating them with 15 instead of 25 radial bins for which we estimate the covariance using 50 rather than 100 jackknife patches, which yields consistent results.

Different measures of detection significance can be defined as follows:
\begin{enumerate}
\item SNR of shear. A simple measure is the tangential shear at the first angular bin outside the trough radius $\theta_T$ in units of its standard deviation according to the jackknife estimate, $g_{\rm t}/\sigma_{\rm g}:=|g_{\rm t}(\theta_{\rm T})|/\sigma_{g_{\rm t}(\theta_{\rm T})}$.
\item For optimal signal-to-noise \citep[cf., e.g.][their Eqn.~11]{2011MNRAS.416.1392G}, we define a linear combination of tangential shear measurements. The weights of the linear combination are chosen as $\bm{W}\propto \hat{\bm{\mathrm{C}}}^{-1}\bm{g_{\rm t}}^{\rm model}$, where we use the model prediction $\bm{g_{\rm t}}^{\rm model}$ for our fiducial bias of $b=1.6$. The SNR of $\Gamma=\bm{W}\cdot\bm{g_{\rm t}}$ is given as $\Gamma/\sigma_{\Gamma}=\Gamma/\sqrt{\bm{W}^{\rm T}\cdot\hat{\bm{\mathrm{C}}}\cdot\bm{W}}$.
\item We do not list a significance based on the $\chi^2$ of the null hypothesis here for two reasons: (I) since the signal is consistent with zero on a range of small and large scales, as is also expected from the model, the $p$-values of the measured $\chi^2$ strongly depend on which bins are used and (II) $\chi^2$ yields an uncertain estimate of SNR due to the variance $\mathrm{Var}(\chi^2)=2n_{\rm bins}$ for $n_{\rm bins}$ bins.
\end{enumerate}
We list these metrics for various trough selections in Table~\ref{tbl:significance}. For the most conservative metric, $g_{\rm t}/\sigma_{\rm g}$, we find a detection significance of $10\sigma$ for the smallest troughs. The optimal linear combination of observables yields even higher significances. Our detection of radial shear around underdensities on small scales of $\theta_{\rm T}=5,10$~arcmin is of considerably higher significance than that of the most recent void lensing studies \citep{2014MNRAS.440.2922M,2014arXiv1404.1834C}. On larger scales, significance decreases but is still comparable.

\subsection{Trough-galaxy angular correlation}
\label{sec:w}

\begin{figure*}
\centering
\includegraphics[width=0.48\textwidth]{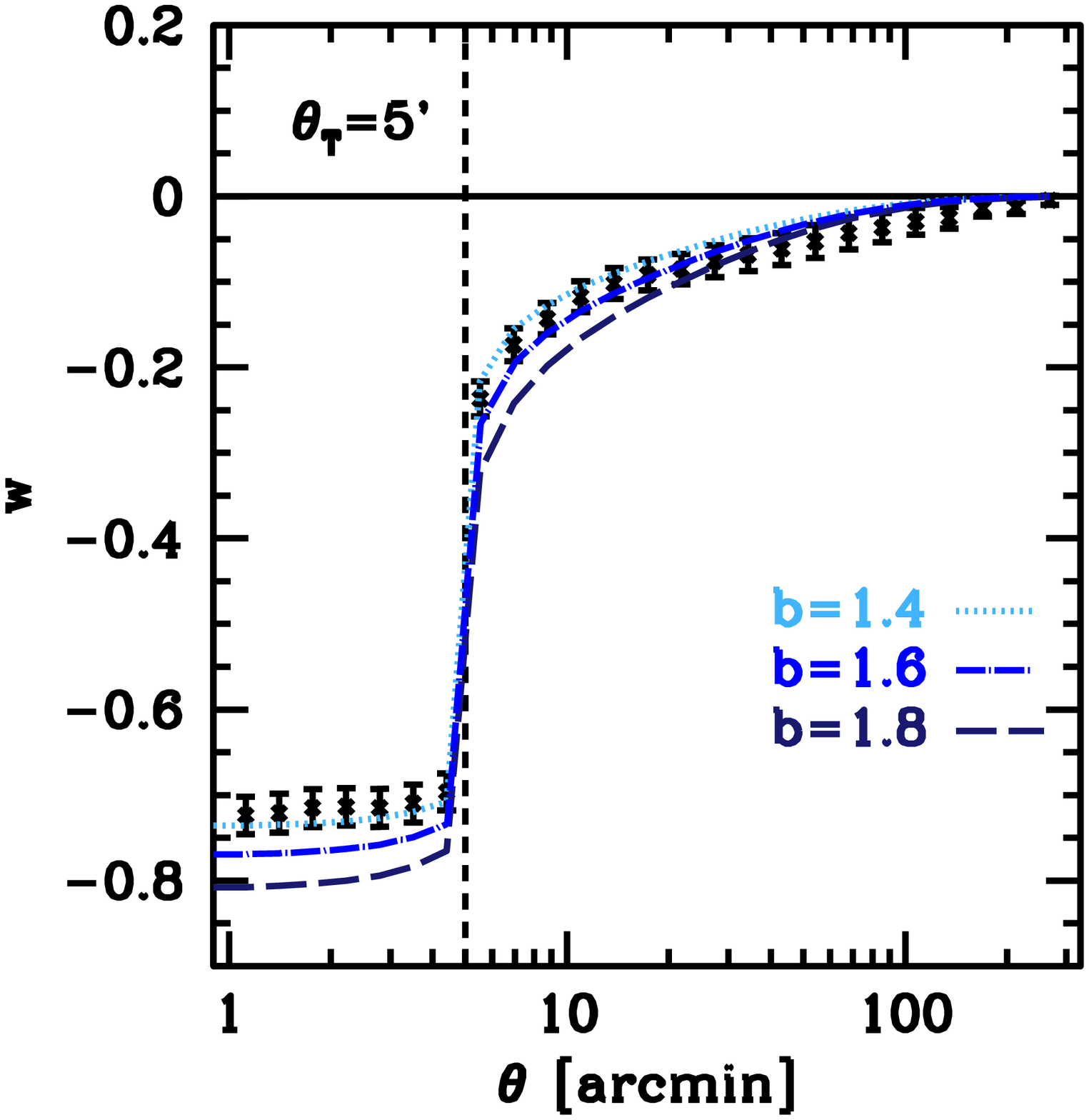}
\includegraphics[width=0.48\textwidth]{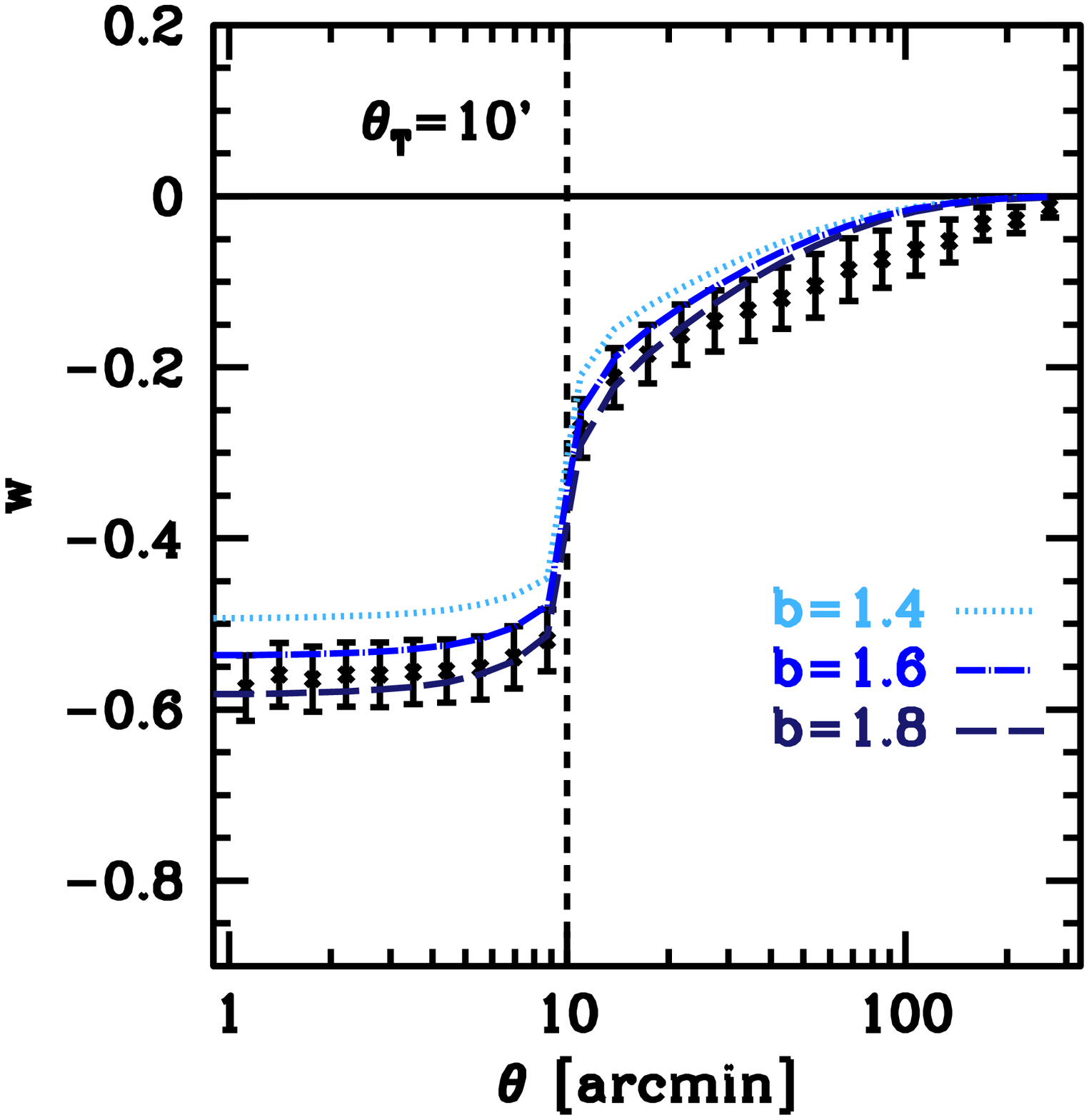}
\includegraphics[width=0.48\textwidth]{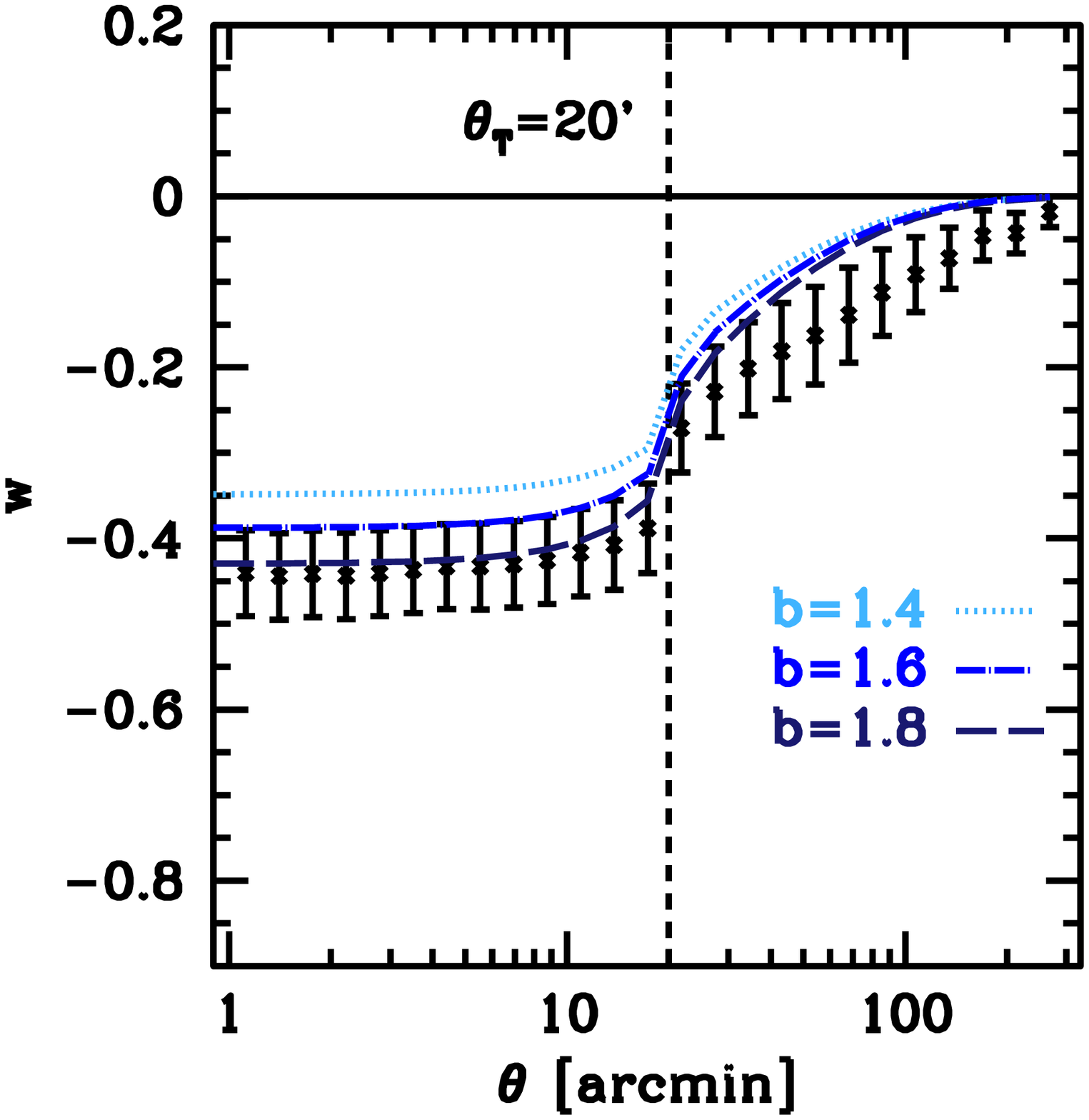}
\includegraphics[width=0.48\textwidth]{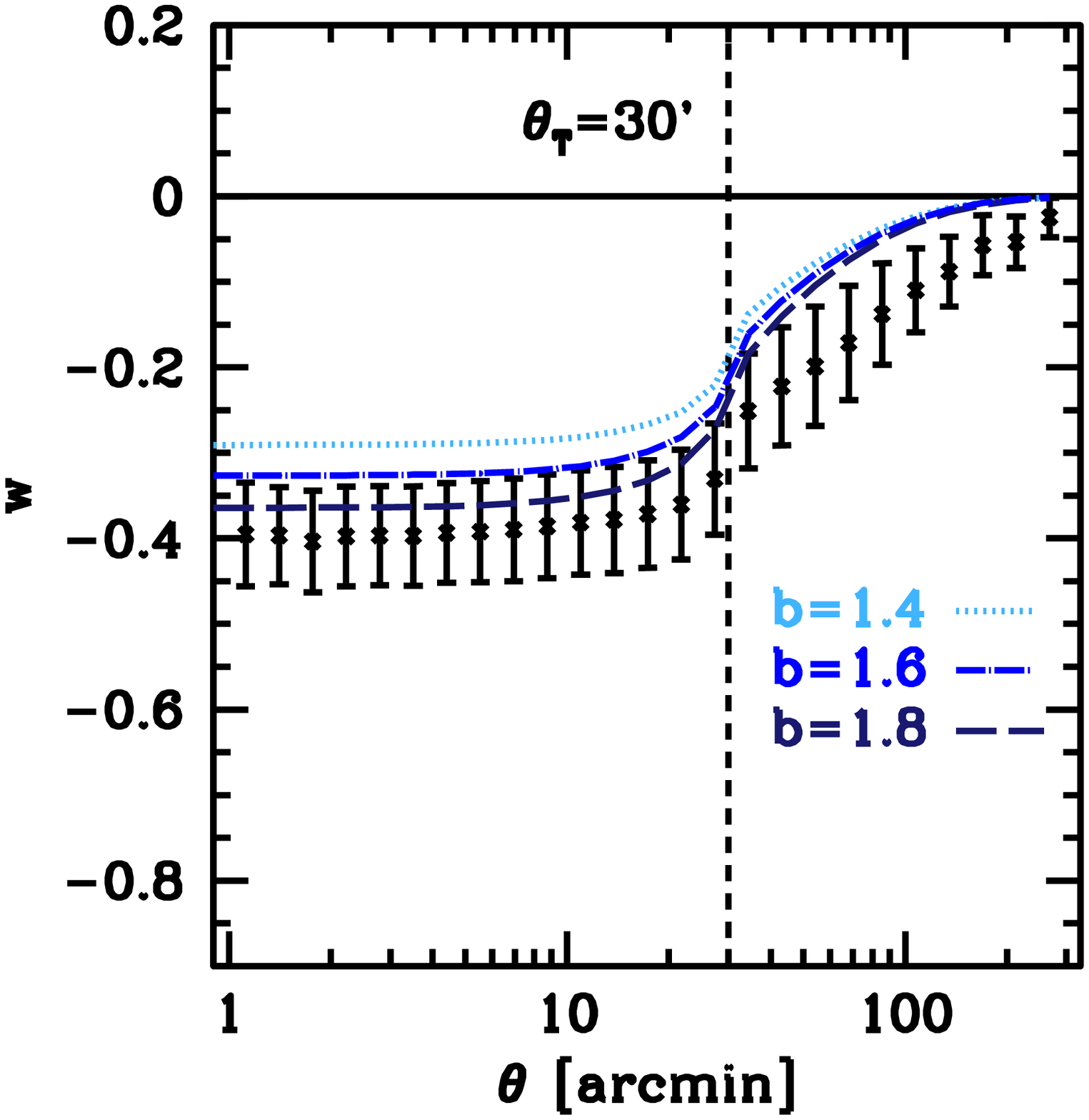}
\caption{Angular two-point correlation of trough positions and redMaGiC galaxies in $z=0.2\ldots0.5$ for the same configurations as in Fig.~\ref{fig:g}. Shown are signal (black) and model predictions, for illustration for different values of the bias ($b=1.4,1.6,1.8$ from light to dark blue, dotted, dot-dashed and dashed lines, cf.~Section~\ref{sec:model}).}
\label{fig:w}
\end{figure*}

The lensing signal around troughs, studied in the previous sections, measures a weighted, projected version of the matter density field (cf.~equation~\ref{eqn:kappadef}). Galaxies themselves also trace the matter field, yet with a different weighting. We have approximated the connection of galaxies to the matter density, so far, as being constant comoving density, deterministic, biased tracers. Measurements of the two-point correlation of trough positions with galaxies are complementary to trough lensing, sensitive to both the properties of the matter field and the details of the connection of galaxies and matter.

We measure the angular two-point correlation between trough positions and redMaGiC galaxies in the same redshift range of $z=0.2\ldots0.5$ and limited to the same survey subarea of 139~deg$^2$ also used for the lensing analysis. Uncertainties are again estimated by estimating the two-point correlation in 100 jackknife resamplings and are highly correlated between bins, as is common in clustering analyses.

Fig.~\ref{fig:w} shows results for the fiducial trough parameters, i.e.~the trough catalogues also used in Fig.~\ref{fig:g}. The low galaxy count level inside the trough, due in part to the selection of regions of low matter density and to Poisson noise, steeply rises at the trough radius outside of which there is no Poisson contribution. Physical, smaller underdensities in the galaxy field are observed out to large radii. Section~\ref{sec:modelw} discusses our modelling of the signal. Although only at moderate significance, there are indications of an increase of bias with trough radius, related to either a general scale or density dependence of bias or assembly bias \citep[e.g.][]{2006ApJ...652...71W}.

\subsection{Comparison to theory}
\label{sec:modelfit}
We briefly compare our measurements to the model put forward in Section~\ref{sec:model}.

Our measurements of tangential shear around underdense troughs are consistent with the predictions at all scales and source and trough redshift configurations tested here. The reduced $\chi^2_{\rm mod}$ of the residual of the data with respect to the $b=1.6$ model are listed in Table~\ref{tbl:significance} and consistent with noise. It is worth noting that the model is a good fit essentially without any free parameter. The only exception to this is the mild dependence on the assumed galaxy bias for the smallest scale $\theta_{\rm T}=5$~arcmin considered here. This can be understood as an effect of the importance of Poisson noise relative to true variations in the matter density field as traced by the galaxies. On large scales where Poisson noise is subdominant, galaxies are dense tracers of the smoothed matter field. Independent of the details of the galaxy placement model, i.e.~as long as galaxy and matter density are somewhat positively correlated, the selection of some percentile in galaxy count then yields an essentially equivalent selection in matter density.

The model also consistently predicts our measurements for the two-point correlation of troughs and galaxies. The estimation of goodness of fit is strongly affected by the correlation of errors over wide ranges of scales. On the larger scales of $\theta_{\rm T}=20$ and $30$~arcmin and when taking into account the full covariance, the model is a good fit to the data for the full range of bias $b=1.6,\ldots,1.8$ probed. For $\theta_{\rm T}=5$ and $10$~arcmin, the large linear bias model with $b=1.8$ is excluded at $4\sigma$ significance (reduced $\chi^2=4.3$ and $2.1$, respectively, for the 25 data points), while the other models are good fits. Similarly, for the shear around overdense cylinders, the only deviation from the prediction is found for the smallest scale cylinders, where the measured shear around overdense cylinders is somewhat larger than predicted. Both observations indicate that at these scales, the linear bias of our galaxy placement model and/or the assumed Gaussianity of the matter field may be not completely valid.

The combination of these measurements therefore is sensitive to both the details of how galaxies follow matter in low- and high-density environments (e.g. a scale or density dependence of galaxy bias) and on cosmological parameters. These aspects will be studied in more detail in Friedrich et al. (in preparation).

\section{Conclusions}

We have presented the measurement of per-mille level radial gravitational shear of background galaxies and negative two-point correlation of foreground galaxies around underdense cylinders (troughs) in the foreground galaxy field in DES data from the SV period. 

\begin{itemize}
\item Our detection of radial shear around these projected underdense regions (cf.~Section~\ref{sec:g}) is highly significant (above $10\sigma$; cf.~Section~\ref{sec:significance}), on the smallest projected scales and widest projection redshift range considered. This is a much higher significance than has been achieved with present data for the shear signal around three-dimensional voids.
\item We develop a model for the shear profile (cf.~Section~\ref{sec:model}), based on the assumption that galaxies are biased, Poissonian tracers of the Gaussian matter density field. The model predicts the lensing measurements consistently within the present level of uncertainty. It is interesting to note that on sufficiently large scales, the prediction is virtually independent of the details of the galaxy placement model, yet sensitive to cosmological parameters (cf.~Friedrich et al., in preparation).
\item Tomographic measurements that split the source sample or the redshift range used for the selection of troughs show consistent results. We note that the significance of radial shears strongly decreases for smaller trough redshift ranges, due to both the increased noise in galaxy counts and the variation of uncorrelated (overdense) structures along the line of sight in front or behind the trough cylinders.
\item We measure the shear signals around underdense and overdense cylinders in galaxy count at the same percentile thresholds (cf.~Section~\ref{sec:symmetry}). On small scales we find indications for some deviation from our simple model predictions for the high-density regions. On large scales, however, we recover the expected symmetry between radial and tangential shear for both cases.
\item In addition to the shear signal, we measure and model the two-point correlation of galaxies from our tracer population around trough positions (cf.~\ref{sec:w}). While consistent with our prediction on sufficiently large scales, this probe is more sensitive to the details of how galaxies trace the matter and therefore complementary to the shear signal.
\end{itemize}

The statistical power of these measurements will strongly increase as larger data sets become available. We note, in particular, that the final survey area of DES will be $\approx30$ times larger at comparable or even better data quality, allowing very precise measurements of the trough lensing signal. With these better statistics, trough lensing will be a relevant probe of cosmology, not only in the sense of constraining parameters of a $\Lambda$CDM model. Also, the potential lack of screening mechanisms in underdense environments would influence the growth of negative density perturbations, with implications for constraining MG models with these measurements.

On small scales, the details of how galaxies trace matter and the intrinsic distribution of the fields involved are likely to play a significant role for model predictions, and simulations in combination with progress on modelling will be required. Under these prerequisites, trough lensing measurements are a promising tool for probing the connection of galaxies and matter and gravity in the underdense Universe.

\section{Acknowledgements}

We are grateful for the extraordinary contributions of our CTIO colleagues and the DECam Construction, Commissioning and Science Verification
teams in achieving the excellent instrument and telescope conditions that have made this work possible.  The success of this project also 
relies critically on the expertise and dedication of the DES Data Management group.

Funding for the DES Projects has been provided by the U.S. Department of Energy, the U.S. National Science Foundation, the Ministry of Science and Education of Spain, the Science and Technology Facilities Council of the United Kingdom, the Higher Education Funding Council for England, the National Center for Supercomputing Applications at the University of Illinois at Urbana-Champaign, the Kavli Institute of Cosmological Physics at the University of Chicago, the Center for Cosmology and Astro-Particle Physics at the Ohio State University,the Mitchell Institute for Fundamental Physics and Astronomy at Texas A\&M University, Financiadora de Estudos e Projetos, Funda{\c c}{\~a}o Carlos Chagas Filho de Amparo {\`a} Pesquisa do Estado do Rio de Janeiro, Conselho Nacional de Desenvolvimento Cient{\'i}fico e Tecnol{\'o}gico and the Minist{\'e}rio da Ci{\^e}ncia, Tecnologia e Inova{\c c}{\~a}o, the Deutsche Forschungsgemeinschaft and the Collaborating Institutions in the Dark Energy Survey. The DES data management system is supported by the National Science Foundation under Grant Number AST-1138766. The DES participants from Spanish institutions are partially supported by MINECO under grants AYA2012-39559, ESP2013-48274, FPA2013-47986, and Centro de Excelencia Severo Ochoa SEV-2012-0234, some of which include ERDF funds from the European Union.

The Collaborating Institutions are Argonne National Laboratory, the University of California at Santa Cruz, the University of Cambridge, Centro de Investigaciones En{\'e}rgeticas, Medioambientales y Tecnol{\'o}gicas-Madrid, the University of Chicago, University College London, the DES-Brazil Consortium, the University of Edinburgh, the Eidgen{\"o}ssische Technische Hochschule (ETH) Z{\"u}rich, Fermi National Accelerator Laboratory, the University of Illinois at Urbana-Champaign, the Institut de Ci{\`e}ncies de l'Espai (IEEC/CSIC), the Institut de F{\'i}sica d'Altes Energies, Lawrence Berkeley National Laboratory, the Ludwig-Maximilians Universit{\"a}t M{\"u}nchen and the associated Excellence Cluster Universe, the University of Michigan, the National Optical Astronomy Observatory, the University of Nottingham, The Ohio State University, the University of Pennsylvania, the University of Portsmouth, SLAC National Accelerator Laboratory, Stanford University, the University of Sussex, and Texas A\&M University.

This project was supported by SFB-Transregio 33 `The Dark Universe' by the Deutsche Forschungsgemeinschaft (DFG) and the DFG cluster of excellence `Origin and Structure of the Universe'. ES is supported by DOE grant DE-AC02-98CH10886. DG and OF acknowledge helpful discussions with Yan-Chuan Cai, Joseph Clampitt, Stefan Hilbert, Ben Hoyle, Richard Kessler, and Carles Sanchez. Measurements of the shear and angular two-point correlation were made using the tree code \textsc{athena} by Martin Kilbinger \citep{2002A&A...396....1S,2014ascl.soft02026K}.
 
This paper has gone through internal review by the DES collaboration.

\addcontentsline{toc}{chapter}{Bibliography}
\bibliographystyle{mn2e}
\bibliography{literature}

\section*{Appendix A: Conditional Probabilities}

The conditional probability density of $\delta_{\rm T}$, given that $N$ galaxies were found inside the radius $\theta_{\rm T}$, is given by
\begin{align}
\label{eq:app1}
p(\delta_{\rm T} | N) &= \frac{P(N|\delta_{\rm T})\, p(\delta_{\rm T})}{P(N)} \nonumber \\
&= \frac{1}{\mathcal{N}}   \frac{\left(\bar N [1+b\delta_{\rm T}] \right)^N }{N!}e^{-\bar N [1+b\delta_{\rm T}]}   \frac{1}{\sqrt{2\uppi \sigma_{\rm T}^2}}e^{-\frac{\delta_{\rm T}^2}{2\sigma_{\rm T}^2}}\nonumber\ ,\\
\end{align}
where we have made the simple assumptions that galaxies trace matter with a constant bias $b$ and that the variation of galaxy counts around the expectation value is given by the Poisson distribution. The normalisation constant gives the overall probability of finding $N$ galaxies inside $\theta_{\rm T}$,
\begin{equation}
\mathcal{N} = P(N).
\end{equation}
In Appendix B, the trough variance $\sigma^2_{\rm T}$ is derived from the 2D power spectrum of the projected matter contrast.

Note that in order to self-consistently define the biased Poisson model as explained above, equation \ref{eq:app1} can only be valid for $\delta_{\rm T} > -1/b$. Furthermore, one has to assume that
\begin{equation}
P\left(N|\delta_{\rm T} \leq -1/b\right) = 0\ \mathrm{for}\ N>0\ .
\end{equation}
As a consequence one also has
\begin{equation}
p\left(\delta_{\rm T}|N\right) = 0\ \mathrm{for}\ N>0\ ,\ \delta_{\rm T} \leq -1/b\ .
\end{equation}
The case $N=0$, however, is more subtle. Here one has $P(N=0|\delta_{\rm T} \leq -1/b) = 1$, and hence
\begin{equation}
p\left(\delta_{\rm T}|N=0\right) = \frac{p(\delta_{\rm T})}{P(N=0)}\ \mathrm{for}\ \delta_{\rm T} \leq -1/b\ .
\end{equation}
This also has to be considered when the probability $P(N=0)$ is computed.

\section*{Appendix B: Variance and Covariance of Convergence and $\delta_{\rm T}$}
\label{sec:App_covariance}

Let $\delta_i$, $i = 1,2$, be two line of sight projections of the matter density contrast $\delta$, i.e.
\begin{equation}
\delta_i(\bm{\theta}) = \int_0^\infty \mathrm d \chi\ q_{i}(\chi) \ \delta(\chi\bm{\theta},  \chi)\ ,
\end{equation}
$q_i$ being the weights of the projections (cf. \citealt{2001PhR...340..291B}). According to the \citet{1954ApJ...119..655L} approximation, the 2D cross power spectrum of $\delta_1$ and $\delta_2$ is given by
\begin{equation}
C_{1, 2}(\ell) = \int_0^\infty \mathrm d \chi\ \frac{q_1(\chi)\,q_2(\chi)}{\chi^2} \ P_\delta\left(\frac{\ell}{\chi},  \chi\right)\ .
\end{equation}
Here, $\chi$ is the comoving distance and a flat universe was assumed. Let $A_i$ be annuli with minimal radius $\theta_{i, \rm{min}}$ and maximal radius $\theta_{i,\rm{max}}$. The annulus-averaged versions of $\delta_i$ are given by
\begin{equation}
D_i(\boldsymbol \theta) = \frac{\int \mathrm d^2 \bm{\theta}'\ G_i(\boldsymbol \theta-\boldsymbol \theta')\ \delta_i(\boldsymbol \theta')}{\uppi(\theta_{i+1,\rm{max}}^2-\theta_{i,\rm{max}}^2)} \; ,
\end{equation}
where $G_i(\boldsymbol \theta)$ is the top-hat filter corresponding to annulus $A_i$.

$\delta_i(\boldsymbol \theta)$ can be expanded into spherical harmonics as follows:
\begin{equation}
\delta_i(\theta, \phi) = \sum_{\ell, m} a_{i,\ell}^m Y_\ell^m(\theta, \phi)\ .
\end{equation}
If $\delta_i$ is a homogeneous and isotropic random field then the coefficients $a_\ell^m$ satisfy the equation \citep[cf.][]{peebles}
\begin{equation}
\langle a_{i,\ell}^m a_{i,\ell'}^{-m'}\rangle = \delta_{\ell\ell'}\delta_{mm'} C_{i,\ell}\ , 
\end{equation}
where $C_{i,\ell}$ is the 2D power spectrum of $\delta_i$. 

Since the expectation values $\langle \delta_i\rangle$ vanish, the covariance $\langle D_1 D_2\rangle$ can be computed as 
\begin{align}
\langle D_1 D_2\rangle &= \int \mathrm{d}\Omega_1\mathrm{d}\Omega_2\ G_1(\boldsymbol\theta_1)G_2(\boldsymbol\theta_2)  \langle\delta_1(\boldsymbol\theta_1)\delta_2(\boldsymbol\theta_2)\rangle \nonumber \\
&= \sum_{\ell, m} \sum_{\ell', m'} \int \mathrm{d}\Omega_1\mathrm{d}\Omega_2\ G_1(\boldsymbol\theta_1)G_2(\boldsymbol\theta_2)  \langle a_\ell^m a_{\ell'}^{-m'}\rangle \times \nonumber \\
&\;\;\;\;\;\;\;\;\;\;\;\;\;\;\;\;\;\;\;\;\;\;\;\;Y_\ell^m(\boldsymbol\theta_1) Y_{\ell'}^{-m'}(\boldsymbol\theta_2) \nonumber \\
&= \sum_{\ell, m} C_\ell  \int \mathrm{d}\Omega_1\ G_1(\boldsymbol\theta_1)   Y_\ell^m(\boldsymbol\theta_1) \times \nonumber \\ 
&\;\;\;\;\;\;\;\;\;\;\;\;\;\;\;\int \mathrm{d}\Omega_2\  G_2(\boldsymbol\theta_2)  Y_\ell^{-m}(\boldsymbol\theta_2) \nonumber \\
&= \sum_{\ell, m} C_\ell  G_{1, \ell}^{-m}G_{2, \ell}^{m}\ ,
\end{align}
where in the last step we used the relation \citep[see e.g.][]{peebles}
\begin{equation}
f_\ell^m = \int \mathrm{d}\Omega\ f(\boldsymbol\theta) Y_\ell^{-m}(\boldsymbol\theta)\ .
\end{equation}
The annuli and circles we will use as filters are isotropic, $\mathrm{i.e.}$ $G_i(\theta, \phi) = G_i(\theta)$. Hence all coefficients $G_{i, \ell}^{m}$ vanish except for $G_{i, \ell}^{0} =: G_{i, \ell}\ $. These are given by
\begin{align}
G_{i, \ell} &= \int \mathrm{d}\Omega\ G_i(\boldsymbol\theta) Y_\ell^{0}(\boldsymbol\theta) \nonumber \\
&= \mathcal{N}_\ell \int_0^\uppi \mathrm{d}\theta\ \int_0^{2\uppi} \mathrm{d}\phi\ \sin(\theta) G_i(\theta) P_\ell(\cos(\theta)) \nonumber \\&= \frac{2\uppi\mathcal{N}_\ell}{A_i} \int_{\theta_{i,\min}}^{\theta_{i,\max}} \mathrm{d}\theta\ \sin(\theta) P_\ell(\cos(\theta))\nonumber \\&= \frac{2\uppi\mathcal{N}_\ell}{A_i} \int_{\cos\theta_{i, \max}}^{\cos\theta_{i,\min}} \mathrm{d}x\ P_\ell(x)\ .
\end{align}
Here, $P_\ell$ are the Legendre polynomials, $A_i$ is the area of the annulus\footnote{The correct expression for the area is $A_i = 2\uppi(\cos\theta_{i, \min}-\cos\theta_{i, \max})$.} and $\mathcal{N}_\ell$ is a normalization factor given by
\begin{equation}
\mathcal{N}_\ell = \sqrt{\frac{2\ell+1}{4\uppi}}\ .
\end{equation}
The covariance then reads
\begin{equation}
\langle D_1 D_2\rangle = \sum_{\ell} C_\ell \, G_{1, \ell}G_{2, \ell}\ .
\end{equation}
Note that, using the equation
\begin{equation}
P_\ell(x) = \frac{1}{2\ell+1}\frac{\mathrm d}{\mathrm d x}\left(P_{\ell+1}(x)-P_{\ell-1}(x)\right)\ , 
\end{equation}
one can simplify the $G_{i, \ell}$ to 
\begin{equation}
G_{i, \ell} = \frac{2\uppi\mathcal{N}_\ell}{(2\ell+1)A_i} \left[ P_{\ell+1}(x)-P_{\ell-1}(x) \right]_{\cos\theta_{\max}}^{\cos\theta_{\min}}\ .
\end{equation}

The covariance of $K_i$ and $\delta_{\rm T}$ can then be computed by setting $\delta_1 = K_i$ and $\delta_2 = \delta_{\rm T}$. The variance $\sigma_{\rm T}^2$ is found by setting both $\delta_1$ and $\delta_2$ to $\delta_{\rm T}$.

\section*{Affiliations}
$^{1}$Universit\"ats-Sternwarte, Fakult\"at f\"ur Physik, Ludwig-Maximilians Universit\"at M\"unchen, Scheinerstr. 1, D-81679 M\"unchen, Germany\\
$^{2}$Max Planck Institute for Extraterrestrial Physics, Giessenbachstrasse, D-85748 Garching, Germany\\
$^{3}$Department of Physics, ETH Zurich, Wolfgang-Pauli-Strasse 16, CH-8093 Zurich, Switzerland\\
$^{4}$Institute of Cosmology \& Gravitation, University of Portsmouth, Portsmouth, PO1 3FX, UK\\
$^{5}$Institut de F\'{\i}sica d'Altes Energies, Universitat Aut\`onoma de Barcelona, E-08193 Bellaterra, Barcelona, Spain\\
$^{6}$Department of Physics and Astronomy, University of Pennsylvania, Philadelphia, PA 19104, USA\\
$^{7}$Kavli Institute for Particle Astrophysics \& Cosmology, P. O. Box 2450, Stanford University, Stanford, CA 94305, USA\\
$^{8}$Department of Physics, University of Arizona, Tucson, AZ 85721, USA\\
$^{9}$SLAC National Accelerator Laboratory, Menlo Park, CA 94025, USA\\
$^{10}$Brookhaven National Laboratory, Bldg 510, Upton, NY 11973, USA\\
$^{11}$Jodrell Bank Center for Astrophysics, School of Physics and Astronomy, University of Manchester, Oxford Road, Manchester M13 9PL, UK\\
$^{12}$Argonne National Laboratory, 9700 South Cass Avenue, Lemont, IL 60439, USA\\
$^{13}$Cerro Tololo Inter-American Observatory, National Optical Astronomy Observatory, Casilla 603, La Serena, Chile\\
$^{14}$Department of Physics \& Astronomy, University College London, Gower Street, London, WC1E 6BT, UK\\
$^{15}$Department of Physics and Electronics, Rhodes University, PO Box 94, Grahamstown, 6140, South Africa\\
$^{16}$Fermi National Accelerator Laboratory, P. O. Box 500, Batavia, IL 60510, USA\\
$^{17}$Department of Astrophysical Sciences, Princeton University, Peyton Hall, Princeton, NJ 08544, USA\\
$^{18}$Institute of Astronomy, University of Cambridge, Madingley Road, Cambridge CB3 0HA, UK\\
$^{19}$Kavli Institute for Cosmology, University of Cambridge, Madingley Road, Cambridge CB3 0HA, UK\\
$^{20}$Institut de Ci\`encies de l'Espai, IEEC-CSIC, Campus UAB, Carrer de Can Magrans, s/n, E-08193 Bellaterra, Barcelona, Spain\\
$^{21}$Department of Physics, Stanford University, 382 Via Pueblo Mall, Stanford, CA 94305, USA\\
$^{22}$Carnegie Observatories, 813 Santa Barbara St., Pasadena, CA 91101, USA\\
$^{23}$CNRS, UMR 7095, Institut d'Astrophysique de Paris, F-75014 Paris, France\\
$^{24}$Sorbonne Universit\'es, UPMC Univ Paris 06, UMR 7095, Institut d'Astrophysique de Paris, F-75014, Paris, France\\
$^{25}$Laborat\'orio Interinstitucional de e-Astronomia - LIneA, Rua Gal. Jos\'e Cristino 77, Rio de Janeiro, RJ - 20921-400, Brazil\\
$^{26}$Observat\'orio Nacional, Rua Gal. Jos\'e Cristino 77, Rio de Janeiro, RJ - 20921-400, Brazil\\
$^{27}$Department of Astronomy, University of Illinois, 1002 W. Green Street, Urbana, IL 61801, USA\\
$^{28}$National Center for Supercomputing Applications, 1205 West Clark St, Urbana, IL 61801, USA\\
$^{29}$School of Physics and Astronomy, University of Southampton,  Southampton, SO17 1BJ, UK\\
$^{30}$George P. and Cynthia Woods Mitchell Institute for Fundamental Physics and Astronomy, and Department of Physics and Astronomy, Texas A\&M University, College Station, TX 77843,  USA\\
$^{31}$Faculty of Physics, Ludwig-Maximilians University, Scheinerstr. 1, D-81679 Munich, Germany\\
$^{32}$Excellence Cluster Universe, Boltzmannstr.\ 2, D-85748 Garching, Germany\\
$^{33}$Jet Propulsion Laboratory, California Institute of Technology, 4800 Oak Grove Dr., Pasadena, CA 91109, USA\\
$^{34}$Kavli Institute for Cosmological Physics, University of Chicago, Chicago, IL 60637, USA\\
$^{35}$Department of Physics, University of Michigan, Ann Arbor, MI 48109, USA\\
$^{36}$Center for Cosmology and Astro-Particle Physics, The Ohio State University, Columbus, OH 43210, USA\\
$^{37}$Department of Physics, The Ohio State University, Columbus, OH 43210, USA\\
$^{38}$Australian Astronomical Observatory, North Ryde, NSW 2113, Australia\\
$^{39}$Departamento de F\'{\i}sica Matem\'atica,  Instituto de F\'{\i}sica, Universidade de S\~ao Paulo,  CP 66318, CEP 05314-970, S\~ao Paulo, SP,  Brazil\\
$^{40}$Department of Astronomy, The Ohio State University, Columbus, OH 43210, USA\\
$^{41}$Department of Astronomy, University of Michigan, Ann Arbor, MI 48109, USA\\
$^{42}$Instituci\'o Catalana de Recerca i Estudis Avan\c{c}ats, E-08010 Barcelona, Spain\\
$^{43}$Department of Physics and Astronomy, Pevensey Building, University of Sussex, Brighton BN1 9QH, UK\\
$^{44}$Centro de Investigaciones Energ\'{e}ticas, Medioambientales y Tecnol\'{o}gicas (CIEMAT), E-28040 Madrid, Spain\\
$^{45}$Department of Physics, University of Illinois, 1110 W. Green St, Urbana, IL 61801, USA\\
$^{46}$SEPnet, South East Physics Network, (www.sepnet.ac.uk)

\label{lastpage}
\end{document}